\documentclass[useAMS,usenatbib]{mn2e}
\usepackage{graphicx}

\newcommand\kms{$\rm{km\,s^{-1}}$}
\newcommand\HII{H\,{\sc ii}}
\newcommand\subsol{$_{\odot}$}

\title[Methanol maser catalogue: Galactic 
Centre]{The 6-GHz methanol multibeam maser catalogue I: 
Galactic Centre region, longitudes 345$^{\circ}$ to 6$^{\circ}$}

\author[J.L.Caswell et al.]
{J. L. Caswell$^1$\thanks{E-mail:james.caswell@csiro.au}, 
G. A. Fuller$^2$, 
J. A. Green$^{1}$,
A. Avison$^2$, S. L. Breen$^{3,1}$, K. J. Brooks$^1$, 
\newauthor
       M. G. Burton$^4$, A. Chrysostomou$^5$, J. Cox$^6$, P. J.
Diamond$^2$, S. P. Ellingsen$^3$, \newauthor
       M. D. Gray$^2$, M. G. Hoare$^7$, M. R. W. Masheder$^8$, N. M. 
McClure-Griffiths$^1$, \newauthor
       M. R. Pestalozzi$^{11,5}$, C. J. Phillips$^1$, L. Quinn$^2$, M. A.
Thompson$^5$, M. A. Voronkov$^1$, \newauthor
       A. J. Walsh$^{9}$, D. Ward-Thompson$^6$, D. Wong-McSweeney$^2$, J. 
A. Yates$^{10}$ \newauthor
        and R. J. Cohen$^{2}$\thanks{Deceased 2006 November 1.} \\ 
$^{1}$ Australia Telescope
National Facility, CSIRO, PO Box 76, Epping, NSW 2121, Australia; \\
$^{2}$  Jodrell Bank Centre for Astrophysics, Alan Turing Building, 
University of Manchester, Manchester, M13 9PL, UK; \\ 
$^{3}$ School of Mathematics and Physics, University of Tasmania, Private
Bag 37, Hobart, TAS 7001, Australia; \\ $^{4}$ School of Physics,
University of New South Wales, Sydney, NSW 2052, Australia;\\ $^{5}$
Centre for Astrophysics Research, Science and Technology Research Institute, 
University of Hertfordshire, Hatfield, AL10 9AB; \\ 
$^{6}$ Department of Physics and Astronomy, Cardiff
University, 5 The Parade, Cardiff, CF24 3YB, UK; \\ $^{7}$ School of
Physics and Astronomy, University of Leeds, Leeds, LS2 9JT, UK; \\ $^{8}$
Astrophysics Group, Department of Physics, Bristol University, Tyndall
Avenue, Bristol, BS8 1TL, UK;  \\ $^{9}$ School of Engineering and Physical Sciences,
James Cook University, Townsville, QLD 4811, Australia;\\ $^{10}$
University College London, Department of Physics and Astronomy, Gower
Street, London, WC1E 6BT, UK; \\ $^{11}$ G\"oteborgs Universitet
Insitutionen f\"or Fysik, G\"oteborg, Sweden \\}

\date{Accepted XXXX . Received XXX; in original form 2009; }

\pagerange{\pageref{firstpage}--\pageref{lastpage}} \pubyear{XXXX}
\begin{document} \maketitle

\label{firstpage}

\begin{abstract}

We have conducted a Galactic plane survey of methanol masers 
at 6668 MHz using a 7-beam receiver on the Parkes telescope.  Here we 
present results from the first part, which provides sensitive unbiased 
coverage of a large region around the Galactic Centre.  Details are 
given for 183 methanol maser sites in the longitude range 345$^{\circ}$ 
through the 
Galactic Centre to 6$^{\circ}$.   Within 6$^{\circ}$ of the Centre, we 
found 88 maser sites, of which more than half (48) are new discoveries.   
The masers are confined to a narrow Galactic latitude range, indicative 
of many sources at the Galactic Centre distance and 
beyond, and confined to a thin disk population;  there is no high latitude 
population that might be ascribed to the Galactic Bulge.  

Within 2$^{\circ}$ of the Galactic Centre the maser velocities all lie 
between -60 and +77 \kms, a range much smaller than the 540 
\kms\ range observed in CO. 
Elsewhere, the maser with highest positive velocity (+107 \kms) 
occurs, surprisingly, near longitude 355$^{\circ}$ and is probably 
attributable to the Galactic bar.   The maser with the most negative 
velocity (-127 \kms)  is near longitude 346$^{\circ}$, within the 
longitude-velocity locus of the near side of the `3-kpc arm'.  It has the 
most extreme velocity of a clear population of masers associated with the 
near and far sides of the 3-kpc arm.  Closer to the Galactic Centre the 
maser space density is generally low, except 
within 0.25 kpc of the Centre itself, the `Galactic Centre Zone', where 
it is 50 times higher, which is hinted at by the longitude 
distribution, and confirmed by the unusual velocities.

\end{abstract}


\section{Introduction}

Methanol masers at 6668-MHz are widespread in our Galaxy and are the 
second strongest cosmic masers known, surpassed only by water (H$_{2}$O) 
at 22~GHz.  Unlike masers of OH, H$_{2}$O and SiO, they appear to occur 
only in association with massive 
%($\ge$8M$_{\odot}$) 
young stars (Minier et al. 2003; Xu et al. 2008).     
A sensitive methanol survey can thus allow a unique census of massive 
star formation taking place in our Galaxy.  Such a survey then has the 
potential to provide a remarkable probe of the spatial distribution and 
kinematics of these most influential Galactic inhabitants, the young 
massive stars.  

This Galactic mapping will be achieved from astrometry of the masers with 
accuracy sufficient to allow precise 
distance measurements from parallax determinations; the maser site 
systemic velocities then allow an
accompanying measure of the velocity field of the Galaxy at each position.
The technique has already been validated for a handful of masers (Reid et 
al. 2009), and over
the coming years we will eventually be able to use a network comprising
hundreds of masers to fully define the Galactic velocity field. 

Equally important, multi-wavelength studies around the masers (see e.g. 
Purcell et al. 2009) provide an 
opportunity to discover the key ingredients in the Galactic environment 
that are necessary to foster massive star formation, a subject of intense 
debate and current uncertainty.  

The necessary foundation for these objectives is a uniformly sensitive 
survey of the whole Galactic Plane for 6668-MHz methanol masers, and 
the Methanol Multibeam (MMB) survey (Green et al. 2009a) was planned to 
meet these needs.    

For such an extensive survey, there are competing demands for prompt 
release of results, and yet also requiring second or third epoch 
observations to provide confirmation and sub-arcsecond positions.  
Accordingly, as our processing is completed, we are 
releasing the survey in several large portions.  We have already 
completed a survey of the Magellanic Clouds in parallel with the Galactic 
Survey  (Green et al. 2008).  We present here the first portion of the 
Galactic survey, which is of special interest since it includes the 
Galactic Centre and thus allows a new assessment of whether the Centre is 
unusual in its star forming activity compared to the spiral arms.  

We also include remarks on some individual sources that possess 
especially interesting properties, such as large velocity extents, 
extreme intensity variability, or occurrence in compact clusters.

\section{The Methanol Multibeam Survey} 

The detailed plans, strategy, and some sample results from the MMB survey 
have been described by Green et al. (2009a).  
In summary, for the present southern sky observations, we first use the 
Parkes 64-m radio telescope, equipped with a seven-beam receiver, and then 
complement this with accurate position measurements from the Australia 
Telescope Compact Array (ATCA).

%The observing parameters are summarized in Table~\ref{paratable}.

The Galactic latitude coverage is  $\mid b \mid \le 2^{\circ}$.  Outside 
this latitude range, previous surveys (see compilation of Pestalozzi et 
al. 2005) have found only four methanol masers within 60$^{\circ}$ 
longitude of the Galactic Centre.
The complete MMB survey will cover the full Galactic plane, 0$^{\circ}$
$<$ \textit{l} $<$ 360$^{\circ}$.  In this paper we present the results 
for the Galactic longitude range from 345$^{\circ}$ to 6$^{\circ}$.   

Our choice for the velocity coverage of the survey was guided by the 
detected range of Galactic CO emission (Dame, Hartmann \& Thaddeus 2001). 
Our instantaneous observing bandwidth of 4~MHz corresponds to a velocity 
range 
of only 180~km~s$^{-1}$ at 6668~MHz but a velocity coverage wider than 
this is needed for regions near the Galactic Centre.  To achieve full 
velocity coverage, the observing frequency incorporated on-line Doppler 
tracking (applied to the rest frequency of 6668.5192 MHz), and we made 
repeated observations at offset velocities where needed.  All regions 
described here have a 
final coverage of at least 325 \kms, which was increased to 610 \kms\ 
for the region within 
2$^{\circ}$ of the Galactic Centre (see Green et al. 2009a).  

The survey sensitivity achieved is a typical rms noise level of 0.17 Jy, 
deeper than all previous unbiased surveys except for the northern region 
of Galactic plane visible to Arecibo (Pandian, Goldsmith \& Deshpande 
2007).

The earlier single dish surveys in the southern hemisphere have already 
been followed up with the ATCA (Caswell 2009 and references therein) to 
determine precise positions, and to determine the range of emission 
emanating from each site within a cluster.  
The efficient observing strategy used by Caswell (1997, 2009) to observe 
large numbers of sources can provide positional accuracy better than 0.4 
arcsec rms.  
With a similar strategy, we have used the ATCA to refine 
the positions of the newly detected masers of the MMB survey.  Likewise, 
we have made new ATCA observations of any previously known sources lacking 
precise positions so as to 
achieve our target accuracy of 0.4 arcsec rms for all maser sites in the 
final survey.  Since the total extent of the multiple spots at a maser site 
is often as large as 1~arcsec, this accuracy is adequate for clear 
identification with associated Galactic objects of other types.  
For our purposes, an efficient strategy with this accuracy is most 
suitable, rather than one achieving an accuracy of 0.1 arcsec rms 
that, in the case of the ATCA, would 
require much longer observing times  (employing individual calibrators 
closer to each target and observed more frequently, longer integration 
times, increased and more uniform hour angle coverage, and restriction to 
good weather conditions).  
 
The spectra of nearly all sources from the survey have been 
re-measured, achieving lower noise than the survey observations, using 
pointed observations at Parkes in `MX' mode (Green et al. 2009a).  They 
have a typical rms noise level of 0.07 Jy and allow 
recognition of new weak features that in some cases 
reveal a velocity range of emission much wider than at first apparent.  
In the case of 5.885-0.393, both survey cube and MX spectra are shown, 
illustrating the lower noise level of the MX spectrum, and variability of 
the maser.  We discuss in Section 4.2 the valuable information on source 
variability acquired through repeat observations.

\section{Survey Results} 

The results are presented in Table 1.  The first 2 columns list the 
Galactic longitude and latitude, commonly used here and elsewhere as a 
source name.  They have been derived from the more precise J2000 
equatorial coordinates given in columns 3 and 4.  We then give velocity 
information, where velocities are relative to the conventional local 
standard of rest (LSR).  Recent suggestions for modifying this LSR are 
discussed in Section 4.6.1.  For the velocity range of the emission 
(lowest and highest velocity of detected emission) we have chosen to list 
the largest range seen at any epoch (see 
Section 4.5).  The velocity of the peak, and the flux density of the peak 
as measured from our high sensitivity follow-up (MX) spectra, are then 
given, followed by a second velocity and peak flux density corresponding 
to the original measurements from the `survey cubes' (SC) themselves;  the 
latter are of lower sensitivity than the `MX' values and can also differ 
because of variability, especially since the epochs differ by as much as 2 
years in some instances (see section 4.2).  Flux densities measured in the 
ATCA data are not given in the table, but in the case of highly variable 
sources, they are discussed in the notes of Section 3.1.  The final column 
of Table 1 gives the epoch of our ATCA measurement or lists
references to positions measured previously which were conducted and 
processed in similar fashion.  
Where the position has been re-determined 
in the current observations and corroborates the earlier value, the 
reference is enclosed in parentheses.

\begin {table*}

\caption{Methanol maser positions.  Reference abbreviations are: C2009: Caswell (2009); CP2008: Caswell \& Phillips (2008); HW1995: Houghton \& Whiteoak (1995); W98: Walsh et al. (1998).}

%\label{testlabel}

%\centerline

\begin{tabular}{lcrcccrcrl} 
\hline

\multicolumn{1}{c}{Source Name} & \multicolumn{2}{c}{Equatorial Coordinates} & \multicolumn{2}{c}{Velocity range} & \multicolumn{2}{c}{MX data} & \multicolumn{2}{c}{Survey Cube data} & \multicolumn{1}{l}{Refs, epoch}\\

\ (~~~l,~~~~~~~b~~~)	&	RA(2000)	&	Dec(2000)	&	$\rm V_{L}$&$\rm V_{H}$	&  $\rm V_{pk}$(MX)	&  $\rm S_{pk}$(MX)  &  $\rm V_{pk}$(SC) & $\rm S_{pk}$(SC)	& 	\\
%\ (~~~$^\circ$~~~~~~~$^\circ$~~~) & (h~~m~~~s) & (~$^\circ$~~ '~~~~") & (\kms) & (\kms) & (\kms) &  (Jy) & (\kms) & (Jy) \\

\ (~~~$^\circ$~~~~~~~$^\circ$~~~) & (h~~m~~~s) & (~$^\circ$~~ '~~~~") & \multicolumn{2}{c}{(\kms)} & (\kms) &  (Jy) & (\kms) & (Jy) \\

\hline

345.003\(-\)0.223	&	17 05 10.89	&	 -41 29 06.2	&	-25.0	&	-20.1	&	-23.1	& 236\hphantom{.12}	&	-23.1	& 193\hphantom{.12}	&	C2009; (2007jul19)	\\
345.003\(-\)0.224	&	17 05 11.23	&	 -41 29 06.9	&	-33.0	&	-25.0	&	-26.2	& 102\hphantom{.12}	&	-26.2	&	85.00	&		C2009	\\
345.010+1.792	&	16 56 47.58	&	 -40 14 25.8	&	-24.0	&	-16.0	&	-21.0	&	268\hphantom{.12}	&	-22.6	& 243\hphantom{.12}	&	C2009	\\
345.012+1.797	&	16 56 46.82	&	 -40 14 08.9	&	-16.0	&	-10.0	&	-12.2	&	34.00	&	-12.2	&	34.00	&				C2009	\\
345.131\(-\)0.174	&	17 05 23.24	&	 -41 21 10.9	&	-31.0	&	-28.0	&	-28.9	&	3.10	&	-28.9	&	4.38	&		2007jul19	\\
345.198\(-\)0.030	&	17 04 59.49	&	 -41 12 45.7	&	-4.0	&	1.0	&	-0.5	&	2.53	&	-0.6	&	2.23	&		2007jul18	\\
345.205+0.317	&	17 03 32.87	&	 -40 59 46.6	&	-64.1	&	-59.9	&	-63.5	&	0.80	&	-60.5	&	1.27	&			2007jul19	\\
345.407\(-\)0.952	&	17 09 35.42	&	 -41 35 57.1	&	-15.5	&	-14.0	&	-14.3	&	2.00	&	-14.3	&	2.10	&			C2009	\\
345.424\(-\)0.951	&	17 09 38.56	&	 -41 35 04.6	&	-21.0	&	-5.0	&	-13.2	&	2.92	&	-13.2	&	4.64	&			C2009	\\
345.441+0.205	&	17 04 46.87	&	 -40 52 38.0	&	-13.0	&	2.0	&	0.9	&	2.27	&	0.9	&	2.32	&			2007jul18	\\
345.487+0.314	&	17 04 28.24	&	 -40 46 28.7	&	-24.0	&	-21.5	&	-22.6	&	2.50	&	-22.6	&	2.50	&			C2009	\\
345.505+0.348	&	17 04 22.91	&	 -40 44 21.7	&	-23.1	&	-10.5	&	-17.8	&	300\hphantom{.12}	&	-17.8	& 307\hphantom{.12}	&	C2009	\\
345.498+1.467	&	16 59 42.84	&	 -40 03 36.1	&	-15.0	&	-13.2	&	-14.2	&	1.20	&	-13.8	&	1.02	&				C2009	\\
345.576\(-\)0.225	&	17 07 01.50	&	 -41 01 43.4	&	-127.2	&	-122.0	&	-126.8	&	0.64	&	-126.8	&	0.65	&		2007jul22	\\
345.807\(-\)0.044	&	17 06 59.85	&	 -40 44 08.2	&	-3.0	&	-0.5	&	-2.0	&	1.00	&	-2.0	&	1.21	&		2007jul18	\\
345.824+0.044	&	17 06 40.70	&	 -40 40 09.9	&	-12.0	&	-9.0	&	-10.3	&	3.17	&	-10.3	&	3.92	&			2007jul18	\\
345.949\(-\)0.268	&	17 08 23.64	&	 -40 45 21.5	&	-22.5	&	-21.4	&	-21.9	&	1.53	&	-21.9	&	1.51	&		2007jul19	\\
345.985\(-\)0.020	&	17 07 27.58	&	 -40 34 43.6	&	-85.5	&	-81.7	&	-83.2	&	5.70	&	-84.1	&	1.41	&		2007jul22	\\
346.036+0.048	&	17 07 20.02	&	 -40 29 49.0	&	-14.5	&	-3.9	&	-6.4	&	8.99	&	-6.4	&	10.42	&			2007jul19	\\
346.231+0.119	&	17 07 39.09	&	 -40 17 53.2	&	-96.6	&	-92.6	&	-95.0	&	1.50	&	-95.0	&	1.37	&			2008aug23	\\
346.480+0.221	&	17 08 00.11	&	 -40 02 15.9	&	-21.0	&	-14.0	&	-18.9	&	30.15	&	-18.9	&	32.02	&				C2009	\\
346.481+0.132	&	17 08 22.72	&	 -40 05 25.6	&	-11.6	&	-4.9	&	-5.5	&	2.10	&	-5.6	&	1.48	&				C2009	\\
346.517+0.117	&	17 08 33.20	&	 -40 04 14.3	&	-3.0	&	1.0	&	-1.7	&	0.30	&	-1.7	&   $<$ 0.30	&				C2009	\\
346.522+0.085	&	17 08 42.29	&	 -40 05 07.8	&	4.7	&	6.1	&	5.7	&	1.90	&	5.7	&	1.47	&				C2009	\\
347.230+0.016	&	17 11 11.18	&	 -39 33 27.2	&	-69.9	&	-68.0	&	-68.9	&	0.86	&	-68.9	&	1.19	&			2007jul22	\\
347.583+0.213	&	17 11 26.72	&	 -39 09 22.5	&	-103.8	&	-96.0	&	-102.3	&	3.18	&	-102.5	&	3.18	&				C2009	\\
347.628+0.149	&	17 11 50.92	&	 -39 09 29.2	&	-98.9	&	-95.0	&	-96.5	&	19.20	&	-96.5	&	18.98	&				C2009	\\
347.631+0.211	&	17 11 36.05	&	 -39 07 07.0	&	-94.0	&	-89.0	&	-91.9	&	5.81	&	-91.9	&	7.17	&				C2009	\\
347.817+0.018	&	17 12 58.05	&	 -39 04 56.1	&	-26.0	&	-22.8	&	-24.1	&	2.52	&	-24.0	&	2.85	&				C2009	\\
347.863+0.019	&	17 13 06.23	&	 -39 02 40.0	&	-37.8	&	-28.0	&	-34.7	&	6.38	&	-34.8	&	6.40	&				C2009	\\
347.902+0.052	&	17 13 05.11	&	 -38 59 35.5	&	-31.5	&	-27.0	&	-27.4	&	5.37	&	-27.5	&	5.46	&				C2009	\\
348.027+0.106	&	17 13 14.12	&	 -38 51 38.8	&	-122.8	&	-114.3	&	-121.2	&	3.07	&	-121.3	&	4.54	&			2007jul22	\\
348.195+0.768	&	17 11 00.20	&	 -38 20 05.5	&	-2.8	&	-0.2	&	-0.8	&	4.55	&	-0.8	&	4.86	&			2007jul18	\\
348.550\(-\)0.979	&	17 19 20.41	&	 -39 03 51.6	&	-19.0	&	-7.0	&	-10.6	&	41.10	&	-10.6	&	36.44	&			C2009	\\
348.550\(-\)0.979n	&	17 19 20.45	&	 -39 03 49.4	&	-23.0	&	-14.0	&	-20.0	&	22.60	&	-20.0	&	18.60	&			C2009	\\
348.579\(-\)0.920	&	17 19 10.61	&	 -39 00 24.2	&	-16.0	&	-14.0	&	-15.1	&	0.32	&	-15.0	&   $<$ 0.30	&			C2009	\\
348.617\(-\)1.162	&	17 20 18.65	&	 -39 06 50.8	&	-21.5	&	-8.5	&	-11.4	&	47.59	&	-11.4	&	44.26	&		2007jul18	\\
348.654+0.244	&	17 14 32.37	&	 -38 16 16.8	&	16.5	&	17.5	&	16.9	&	0.82	&	16.9	&	0.99	&			2007jul18	\\
348.723\(-\)0.078	&	17 16 04.77	&	 -38 24 08.8	&	9.0	&	12.0	&	11.5	&	2.58	&	11.5	&	2.25	&		2007jul18	\\
348.703\(-\)1.043	&	17 20 04.06	&	 -38 58 30.9	&	-17.5	&	-2.5	&	-3.5	&	65.00	&	-3.5	&	62.00	&			C2009	\\
348.727\(-\)1.037	&	17 20 06.54	&	 -38 57 09.1	&	-12.0	&	-6.0	&	-7.4	&	80.78	&	-7.4	&	72.85	&			C2009	\\
348.884+0.096	&	17 15 50.13	&	 -38 10 12.4	&	-79.0	&	-73.0	&	-74.5	&	12.18	&	-74.5	&	12.86	&				C2009	\\
348.892\(-\)0.180	&	17 17 00.23	&	 -38 19 28.9	&	1.0	&	2.0	&	1.5	&	2.70	&	1.5	&	2.44	&			C2009	\\
349.067\(-\)0.017	&	17 16 50.74	&	 -38 05 14.3	&	6.0	&	16.0	&	11.6	&	2.30	&	11.6	&	2.43	&			C2009	\\
349.092+0.105	&	17 16 24.74	&	 -37 59 47.2	&	-78.0	&	-74.0	&	-76.6	&	33.30	&	-76.5	&	23.09	&				C2009	\\
349.092+0.106	&	17 16 24.59	&	 -37 59 45.8	&	-83.0	&	-78.0	&	-81.5	&	9.90	&	-81.5	&	10.40	&				C2009	\\
349.151+0.021	&	17 16 55.88	&	 -37 59 47.9	&	14.1	&	25.0	&	14.6	&	3.36	&	14.6	&	3.33	&			2007jul18	\\
349.579\(-\)0.679	&	17 21 05.44	&	 -38 02 54.7	&	-26.0	&	-24.0	&	-25.0	&	1.90	&	-25.0	&	5.86	&		2007jul19	\\
349.799+0.108	&	17 18 27.74	&	 -37 25 03.5	&	-65.5	&	-57.4	&	-64.7	&	3.00	&	-62.4	&	2.11	&			2007jul22	\\
349.884+0.231	&	17 18 12.37	&	 -37 16 40.0	&	13.5	&	17.5	&	16.2	&	6.96	&	16.2	&	6.42	&			2007jul18	\\
350.011\(-\)1.342	&	17 25 06.54	&	 -38 04 00.7	&	-28.0	&	-25.0	&	-25.8	&	2.38	&	-25.8	&	2.02	&			C2009	\\
350.015+0.433	&	17 17 45.45	&	 -37 03 11.9	&	-37.0	&	-29.0	&	-30.3	&	7.20	&	-30.4	&	9.02	&				C2009	\\
350.104+0.084	&	17 19 26.68	&	 -37 10 53.1	&	-69.0	&	-67.5	&	-68.1	&	9.90	&	-68.1	&	14.60	&				C2009	\\
350.105+0.083	&	17 19 27.01	&	 -37 10 53.3	&	-76.0	&	-61.0	&	-74.1	&	13.60	&	-74.1	&	15.21	&				C2009	\\
350.116+0.084	&	17 19 28.83	&	 -37 10 18.8	&	-69.0	&	-67.0	&	-68.0	&	10.30	&	-68.0	&	9.50	&				C2009	\\
350.116+0.220	&	17 18 55.11	&	 -37 05 38.1	&	3.0	&	5.0	&	4.2	&	2.78	&	4.2	&	2.31	&			2007jul18	\\
350.189+0.003	&	17 20 01.41	&	 -37 09 30.7	&	-65.0	&	-62.0	&	-62.4	&	1.07	&	-62.4	&	1.30	&			2007nov26	\\
350.299+0.122	&	17 19 50.87	&	 -36 59 59.9	&	-70.0	&	-61.0	&	-62.1	&	31.34	&	-62.2	&	31.17	&		C2009; (2007jul19)	\\
350.340+0.141	&	17 19 53.43	&	 -36 57 18.8	&	-60.0	&	-57.5	&	-58.4	&	2.50	&	-58.4	&	2.40	&			2007jul19	\\
350.344+0.116	&	17 20 00.03	&	 -36 58 00.1	&	-66.0	&	-55.0	&	-65.4	&	19.90	&	-65.4	&	18.90	&		C2009; (2007jul19)	\\
350.356\(-\)0.068	&	17 20 47.55	&	 -37 03 42.0	&	-68.5	&	-66.0	&	-67.6	&	1.44	&	-67.6	&	1.40	&		2007jul22	\\

\end{tabular}
\end{table*}

\begin {table*}

\addtocounter{table}{-1}

%\caption{ - continued p2 of 3}
\caption{\textit{- continued p2 of 3}}

%\centerline

\begin{tabular}{lcrcccrcrl} 

\hline

\multicolumn{1}{c}{Source Name} & \multicolumn{2}{c}{Equatorial Coordinates} & \multicolumn{2}{c}{Velocity range} & \multicolumn{2}{c}{MX data} & \multicolumn{2}{c}{Survey Cube data} & \multicolumn{1}{l}{Refs, epoch}\\

\ (~~~l,~~~~~~~b~~~)	&	RA(2000)	&	Dec(2000)	&	$\rm V_{L}$&$\rm V_{H}$	&  $\rm V_{pk}$(MX)	&  $\rm S_{pk}$(MX)  &  $\rm V_{pk}$(SC) & $\rm S_{pk}$(SC)	& 	\\

\ (~~~$^\circ$~~~~~~~$^\circ$~~~) & (h~~m~~~s) & (~$^\circ$~~ '~~~~") & \multicolumn{2}{c}{(\kms)} & (\kms) &  (Jy) & (\kms) & (Jy) \\

\hline

350.470+0.029	&	17 20 43.24	&	 -36 54 46.6	&	-11.0	&	-5.5	&	-6.3	&	1.44	&	-6.3	&	1.00	&			2007nov26	\\
350.520\(-\)0.350	&	17 22 25.32	&	 -37 05 13.4	&	-25.0	&	-22.0	&	-24.6	&	1.67	&	-24.6	&	1.04	&		2007jul19	\\
350.686\(-\)0.491	&	17 23 28.63	&	 -37 01 48.8	&	-15.0	&	-13.0	&	-13.7	&	17.85	&	-13.8	&	19.38	&			C2009	\\
350.776+0.138	&	17 21 08.58	&	 -36 35 58.8	&	34.5	&	39.0	&	38.7	&	0.65	&	38.7	&	0.91	&			2007jul21	\\
351.161+0.697	&	17 19 57.50	&	 -35 57 52.8	&	-7.0	&	-2.0	&	-5.2	&	17.02	&	-5.2	&	12.02	&				C2009	\\
351.242+0.670	&	17 20 17.84	&	 -35 54 46.0	&	2.0	&	3.0	&	2.5	&	0.74	&	2.5	&	2.40	&			2007jul21; (CP2008)	\\
351.251+0.652	&	17 20 23.87	&	 -35 54 57.0	&	-7.5	&	-6.0	&	-7.1	&	0.99	&	-7.1	&	1.30	&			2007jul21	\\
351.382\(-\)0.181	&	17 24 09.58	&	 -36 16 49.3	&	-69.0	&	-58.0	&	-59.7	&	19.66	&	-59.8	&	16.85	&		2007jul22	\\
351.417+0.645	&	17 20 53.37	&	 -35 47 01.2	&	-12.0	&	-6.0	&	-10.4	& 3423\hphantom{.12}	&	-10.4	& 3506\hphantom{.12}	&		C2009	\\
351.417+0.646	&	17 20 53.18	&	 -35 46 59.3	&	-12.0	&	-7.0	&	-11.1	& 1840\hphantom{.12}	&	-11.2	&	1816\hphantom{.12}	&	C2009	\\
351.445+0.660	&	17 20 54.61	&	 -35 45 08.6	&	-14.0	&	1.0	&	-7.1	& 129\hphantom{.12}	&	-7.1	&	110\hphantom{.12}	&	C2009	\\
351.581\(-\)0.353	&	17 25 25.12	&	 -36 12 46.1	&	-100.0	&	-88.0	&	-94.2	&	47.50	&	-94.2	&	47.46	&			C2009	\\
351.611+0.172	&	17 23 21.25	&	 -35 53 32.6	&	-46.0	&	-31.5	&	-43.7	&	4.2	&	-43.7	&	4.1	&			2007jul19	\\
351.688+0.171	&	17 23 34.52	&	 -35 49 46.3	&	-47.5	&	-35.0	&	-36.1	&	41.54	&	-36.1	&	42.14	&			2007jul19	\\
351.775\(-\)0.536	&	17 26 42.57	&	 -36 09 17.6	&	-9.0	&	3.0	&	1.3	& 231\hphantom{.12}	&	1.3	& 311\hphantom{.12}	&	C2009	\\
352.083+0.167	&	17 24 41.22	&	 -35 30 18.6	&	-68.2	&	-63.6	&	-66.0	&	6.77	&	-66.0	&	5.77	&				C2009	\\
352.111+0.176	&	17 24 43.56	&	 -35 28 38.4	&	-61.0	&	-50.0	&	-54.8	&	7.46	&	-54.8	&	5.93	&				C2009	\\
352.133\(-\)0.944	&	17 29 22.32	&	 -36 05 00.2	&	-18.8	&	-5.6	&	-7.7	&	16.32	&	-7.8	&	15.66	&			C2009	\\
352.517\(-\)0.155	&	17 27 11.34	&	 -35 19 32.4	&	-52.0	&	-49.0	&	-51.2	&	9.69	&	-51.3	&	9.42	&		2007jul22; (C2009)	\\
352.525\(-\)0.158	&	17 27 13.42	&	 -35 19 15.5	&	-62.0	&	-52.0	&	-53.0	&	0.70	&	-53.0	&	0.70	&			C2009	\\
352.584\(-\)0.185	&	17 27 29.58	&	 -35 17 14.6	&	-92.6	&	-79.7	&	-85.7	&	6.38	&	-85.6	&	3.73	&		2007jul22	\\
352.604\(-\)0.225	&	17 27 42.73	&	 -35 17 34.2	&	-85.0	&	-81.0	&	-81.7	&	3.30	&	-81.8	&	1.80	&		2007may22	\\
352.624\(-\)1.077	&	17 31 15.31	&	 -35 44 47.7	&	-2.0	&	7.0	&	5.8	&	20.00	&	5.8	&	21.00	&			C2009	\\
352.630\(-\)1.067	&	17 31 13.91	&	 -35 44 08.7	&	-8.0	&	-2.0	&	-2.9	& 183\hphantom{.12}	&	-3.0	& 137\hphantom{.12}	&	C2009	\\
352.855\(-\)0.201	&	17 28 17.59	&	 -35 04 12.9	&	-54.1	&	-50.1	&	-51.3	&	1.29	&	-51.4	&	1.45	&		2007jul19	\\
353.216\(-\)0.249	&	17 29 27.80	&	 -34 47 47.3	&	-25.0	&	-15.0	&	-22.9	&	5.14	&	-23.0	&	1.10	&		2007jul19	\\
353.273+0.641	&	17 26 01.58	&	 -34 15 15.4	&	-7.0	&	-3.0	&	-4.4	&	8.30	&	-4.4	&	12.70	&			2007jul21; (C2009)	\\
353.363\(-\)0.166	&	17 29 31.40	&	 -34 37 40.3	&	-80.1	&	-78.3	&	-79.0	&	2.79	&	-79.0	&	2.94	&		2007jul22	\\
353.370\(-\)0.091	&	17 29 14.27	&	 -34 34 50.2	&	-56.0	&	-43.4	&	-44.7	&	1.35	&	-45.7	&	1.13	&		2007jul19	\\
353.378+0.438	&	17 27 07.59	&	 -34 16 50.5	&	-16.5	&	-14.0	&	-15.7	&	0.97	&	-15.7	&	1.06	&			2007nov26	\\
353.410\(-\)0.360	&	17 30 26.18	&	 -34 41 45.6	&	-23.0	&	-19.0	&	-20.3	& 116\hphantom{.12}	&	-20.4	& 109\hphantom{.12}	&	C2009	\\
353.429\(-\)0.090	&	17 29 23.48	&	 -34 31 50.3	&	-63.9	&	-45.0	&	-61.8	&	13.39	&	-61.8	&	13.00	&		2007jul22	\\
353.464+0.562	&	17 26 51.53	&	 -34 08 25.7	&	-52.7	&	-48.7	&	-50.3	&	11.88	&	-50.3	&	12.84	&				C2009	\\
353.537\(-\)0.091	&	17 29 41.25	&	 -34 26 28.4	&	-59.0	&	-54.0	&	-56.6	&	2.51	&	-56.6	&	2.31	&		2007jul19	\\
354.206\(-\)0.038	&	17 31 15.01	&	 -33 51 15.1	&	-37.5	&	-35.0	&	-37.1	&	1.11	&	-37.1	&	1.43	&		2007feb05	\\
354.308\(-\)0.110	&	17 31 48.56	&	 -33 48 29.1	&	11.0	&	19.5	&	18.8	&	3.44	&	18.7	&	3.89	&		2007feb05	\\
354.496+0.083	&	17 31 31.77	&	 -33 32 44.0	&	17.5	&	27.5	&	27.0	&	8.41	&	26.9	&	7.78	&			2007feb05	\\
354.615+0.472	&	17 30 17.13	&	 -33 13 55.1	&	-27.0	&	-12.5	&	-24.4	&	166\hphantom{.12}	&	-24.3	& 185\hphantom{.12} &	2007feb05; (C2009)	\\
354.701+0.299	&	17 31 12.06	&	 -33 15 16.7	&	98.0	&	104.0	&	102.8	&	1.29	&	102.7	&	1.20	&			2007feb05	\\
354.724+0.300	&	17 31 15.55	&	 -33 14 05.7	&	91.0	&	95.0	&	93.9	&	12.58	&	93.8	&	12.16	&			2007jul22: (C2009)	\\
355.184\(-\)0.419	&	17 35 20.49	&	 -33 14 28.6	&	-2.0	&	-0.5	&		&		&	-1.4	&	1.35	&		2007feb05	\\
355.343+0.148	&	17 33 28.84	&	 -32 48 00.2	&	4.0	&	7.0	&	5.8	&	1.24	&	5.8	&	1.20	&			2007jul21; (2009)	\\
355.344+0.147	&	17 33 29.06	&	 -32 47 58.9	&	19.0	&	21.0	&	19.9	&	10.17	&	19.9	&	10.49	&			2007jul21; (2009)	\\
355.346+0.149	&	17 33 28.91	&	 -32 47 49.5	&	9.0	&	12.5	&	10.5	&	7.39	&	9.9	&	9.16	&			2007jul21; (2009)	\\
355.538\(-\)0.105	&	17 34 59.60	&	 -32 46 22.7	&	-3.5	&	5.0	&	3.8	&	1.25	&	3.8	&	1.50	&		2007feb05	\\
355.545\(-\)0.103	&	17 35 00.28	&	 -32 45 58.1	&	-31.0	&	-27.5	&	-28.2	&	1.22	&	-28.2	&	1.25	&		2007feb05	\\
355.642+0.398	&	17 33 15.40	&	 -32 24 46.8	&	-9.0	&	-6.9	&	-7.9	&	1.44	&	-7.9	&	1.90	&			2007feb05	\\
355.666+0.374	&	17 33 24.92	&	 -32 24 21.1	&	-4.5	&	0.6	&	-3.3	&	2.47	&	-3.4	&	1.93	&			2007feb05	\\
356.054\(-\)0.095	&	17 36 16.55	&	 -32 20 02.7	&	15.6	&	17.7	&	16.7	&	0.52	&	16.9	&	0.76	&		2007jul21	\\
356.662\(-\)0.263	&	17 38 29.16	&	 -31 54 38.8	&	-57.0	&	-44.0	&	-53.8	&	8.38	&	-53.8	&	10.08	&			C2009	\\
357.558\(-\)0.321	&	17 40 57.19	&	 -31 10 59.3	&	-5.5	&	0.0	&	-3.9	&	2.16	&	-3.9	&	1.95	&		2007jul21; 2006mar31	\\
357.559\(-\)0.321	&	17 40 57.33	&	 -31 10 56.9	&	15.0	&	18.0	&	16.2	&	2.01	&	16.2	&	2.00	&		2007jul21	\\
357.922\(-\)0.337	&	17 41 54.94	&	 -30 52 55.1	&	-5.5	&	-4.0	&	-4.6	&	0.97	&	-4.9	&	1.50	&		2006mar31	\\
357.924\(-\)0.337	&	17 41 55.17	&	 -30 52 50.2	&	-4.5	&	3.0	&	-2.1	&	2.34	&	-2.1	&	3.37	&		2006mar31	\\
357.965\(-\)0.164	&	17 41 20.14	&	 -30 45 14.4	&	-9.0	&	3.0	&	-8.6	&	2.74	&	-8.8	&	3.10	&			C2009	\\
357.967\(-\)0.163	&	17 41 20.26	&	 -30 45 06.9	&	-6.0	&	0.0	&	-3.1	&	47.50	&	-4.2	&	55.14	&			C2009	\\
358.263\(-\)2.061	&	17 49 37.63	&	 -31 29 18.0	&	0.5	&	6.0	&	5.0	&	17.20	&		&		&			C2009	\\
358.371\(-\)0.468	&	17 43 31.95	&	 -30 34 10.7	&	-1.0	&	13.0	&	1.3	&	44.01	&	1.2	&	46.87	&			C2009	\\
358.386\(-\)0.483	&	17 43 37.83	&	 -30 33 51.1	&	-7.0	&	-5.0	&	-6.0	&	6.95	&	-6.0	&	11.60	&			C2009	\\
358.460\(-\)0.391	&	17 43 26.76	&	 -30 27 11.3	&	-0.5	&	4.0	&	1.3	&	47.73	&	1.2	&	25.40	&		2006mar31	\\
358.460\(-\)0.393	&	17 43 27.24	&	 -30 27 14.6	&	-8.5	&	6.0	&	-7.3	&	11.19	&	-7.5	&	14.50	&		2006mar31	\\

\end{tabular}
\end{table*}

\begin {table*}

\addtocounter{table}{-1}

%\caption{ - continued p3 of 3}
\caption{\textit{- continued p3 of 3}}

%\centerline
\begin{tabular}{lcrcccrcrl} 

\hline

\multicolumn{1}{c}{Source Name} & \multicolumn{2}{c}{Equatorial Coordinates} & \multicolumn{2}{c}{Velocity range} & \multicolumn{2}{c}{MX data} & \multicolumn{2}{c}{Survey Cube data} & \multicolumn{1}{l}{Refs, epoch}\\

\ (~~~l,~~~~~~~b~~~)	&	RA(2000)	&	Dec(2000)	&	$\rm V_{L}$&$\rm V_{H}$	&  $\rm V_{pk}$(MX)	&  $\rm S_{pk}$(MX)  &  $\rm V_{pk}$(SC) & $\rm S_{pk}$(SC)	& 	\\

\ (~~~$^\circ$~~~~~~~$^\circ$~~~) & (h~~m~~~s) & (~$^\circ$~~ '~~~~") & \multicolumn{2}{c}{(\kms)} & (\kms) &  (Jy) & (\kms) & (Jy) \\

\hline

358.721\(-\)0.126	&	17 43 02.31	&	 -30 05 29.9	&	8.8	&	13.9	&	10.6	&	2.99	&	10.5	&	3.64	&		2006mar31	\\
358.809\(-\)0.085	&	17 43 05.40	&	 -29 59 45.8	&	-60.3	&	-50.5	&	-56.2	&	6.86	&	-56.2	&	11.99	&		2006mar31	\\
358.841\(-\)0.737	&	17 45 44.29	&	 -30 18 33.6	&	-30.0	&	-17.0	&	-20.7	&	10.94	&	-20.6	&	13.11	&		2006mar31	\\
358.906+0.106	&	17 42 34.57	&	 -29 48 46.8	&	-20.5	&	-16.5	&	-18.1	&	1.70	&	-18.1	&	2.39	&			2006mar31	\\
358.931\(-\)0.030	&	17 43 10.02	&	 -29 51 45.8	&	-22.0	&	-14.5	&	-15.9	&	5.90	&	-15.9	&	10.06	&		2006mar31	\\
358.980+0.084	&	17 42 50.44	&	 -29 45 40.4	&	5.0	&	7.0	&	6.2	&   $<$ 0.20	&	6.2	&	1.6	&			2007nov25	\\
359.138+0.031	&	17 43 25.67	&	 -29 39 17.3	&	-7.0	&	1.0	&	-3.9	&	15.42	&	-3.9	&	19.61	&				C2009	\\
359.436\(-\)0.104	&	17 44 40.60	&	 -29 28 16.0	&	-53.0	&	-45.0	&	-47.8	&	73.50	&	-46.7	&	59.70	&			C2009	\\
359.436\(-\)0.102	&	17 44 40.21	&	 -29 28 12.5	&	-58.0	&	-54.0	&	-53.3	&	1.65	&	-53.6	&	1.50	&			C2009	\\
359.615\(-\)0.243	&	17 45 39.09	&	 -29 23 30.0	&	14.0	&	27.0	&	19.3	&	38.62	&	22.6	&	71.70	&			C2009	\\
359.938+0.170	&	17 44 48.55	&	 -28 53 59.4	&	-1.5	&	0.2	&	-0.5	&	2.34	&	-0.5	&	1.56	&			2007jul21	\\
359.970\(-\)0.457	&	17 47 20.17	&	 -29 11 59.4	&	20.0	&	24.1	&	23.0	&	2.39	&	23.8	&	2.24	&			C2009	\\
0.092+0.663	&	17 48 25.90	&	 -29 12 05.9	&	10.0	&	25.0	&	23.8	&	18.86	&	23.5	&	24.80	&			2006mar31	\\
0.167\(-\)0.446	&	17 47 45.46	&	 -29 01 29.3	&	9.5	&	17.0	&	13.8	&	1.33	&	13.8	&	4.44	&			2006mar31	\\
0.212\(-\)0.001	&	17 46 07.63	&	 -28 45 20.9	&	41.0	&	50.5	&	49.5	&	3.32	&	49.3	&	3.47	&				C2009	\\
0.315\(-\)0.201	&	17 47 09.13	&	 -28 46 15.7	&	14.0	&	27.0	&	19.4	&	62.60	&	19.4	&	72.16	&				C2009	\\
0.316\(-\)0.201	&	17 47 09.33	&	 -28 46 16.0	&	20.0	&	22.0	&	21.1	&	0.58	&	21.0	&  	0.60	&				C2009	\\
0.376+0.040	&	17 46 21.41	&	 -28 35 40.0	&	35.0	&	40.0	&	37.1	&	0.62	&	37.0	&	2.32	&				C2009	\\
0.409\(-\)0.504	&	17 48 33.48	&	 -28 50 52.5	&	24.5	&	27.0	&	25.4	&	2.61	&	25.3	&	2.77	&			2007jul31	\\
0.475\(-\)0.010	&	17 46 47.07	&	 -28 32 06.9	&	23.0	&	31.0	&	28.8	&	3.14	&	28.8	&	3.43	&			2006mar31 (C2009)	\\
0.496+0.188	&	17 46 03.96	&	 -28 24 52.8	&	-12.0	&	2.0	&	0.9	&	24.51	&	0.8	&	32.14	&				C2009	\\
0.546\(-\)0.852	&	17 50 14.35	&	 -28 54 31.1	&	8.0	&	20.0	&	11.8	&	61.92	&	11.8	&	62.83	&				C2009	\\
0.645\(-\)0.042	&	17 47 18.65	&	 -28 24 25.0	&	46.0	&	53.0	&	49.5	&	54.39	&	49.5	&	76.08	&			HW1995	\\
0.647\(-\)0.055	&	17 47 22.04	&	 -28 24 42.6	&	49.0	&	52.0	&		&		&	51.0	&		&			HW1995 (2.0 Jy)	\\
0.651\(-\)0.049	&	17 47 21.12	&	 -28 24 18.2	&	46.0	&	49.0	&	48.3	&	21.45	&	48.0	&	24.00	&			HW1995	\\
0.657\(-\)0.041	&	17 47 20.05	&	 -28 23 46.5	&	48.0	&	56.0	&		&		&	49.9	&		&			HW1995 (1.8 Jy)	\\
0.665\(-\)0.036	&	17 47 20.04	&	 -28 23 12.8	&	58.0	&	62.0	&	60.4	&	2.61	&	60.4	&	6.00	&			HW1995	\\
0.666\(-\)0.029	&	17 47 18.64	&	 -28 22 54.6	&	68.0	&	73.0	&	70.5	&	34.38	&	70.0	&	32.90	&			HW1995	\\
0.667\(-\)0.034	&	17 47 19.87	&	 -28 23 01.3	&	49.0	&	56.0	&		&		&	55.0	&		&			HW1995 (0.4 Jy) \\
0.672\(-\)0.031	&	17 47 20.04	&	 -28 22 41.3	&	55.0	&	59.0	&	58.2	&	7.29	&	58.2	&	9.00	&			HW1995	\\
0.673\(-\)0.029	&	17 47 19.54	&	 -28 22 32.6	&	65.5	&	66.5	&		&		&	66.0	&		&			HW1995 (0.4 Jy)	\\
0.677\(-\)0.025	&	17 47 19.28	&	 -28 22 14.8	&	70.0	&	77.0	&	73.3	&	4.87	&	73.3	&	4.00	&			HW1995	\\
0.695\(-\)0.038	&	17 47 24.74	&	 -28 21 43.6	&	64.0	&	75.0	&	68.6	&	32.33	&	68.6	&	36.41	&			HW1995	\\
0.836+0.184	&	17 46 52.86	&	 -28 07 34.8	&	2.0	&	5.0	&	3.5	&	6.64	&	3.6	&	8.99	&				C2009	\\
1.008\(-\)0.237	&	17 48 55.29	&	 -28 11 47.9	&	1.0	&	7.0	&	1.6	&	13.57	&	1.6	&	15.23	&			2006mar31	\\
1.147\(-\)0.124	&	17 48 48.53	&	 -28 01 11.2	&	-20.5	&	-14.0	&	-15.3	&	3.01	&	-15.3	&	2.97	&			2007jul19; (W98)	\\
1.329+0.150	&	17 48 10.31	&	 -27 43 20.7	&	-13.5	&	-11.0	&	-12.2	&	2.08	&	-12.0	&	1.56	&			2006mar31	\\
1.719\(-\)0.088	&	17 49 59.84	&	 -27 30 36.9	&	-9.0	&	-4.5	&	-8.1	&	7.82	&	-8.0	&	9.81	&			2006mar31	\\
2.143+0.009	&	17 50 36.14	&	 -27 05 46.5	&	54.0	&	65.0	&	62.6	&	7.08	&	62.7	&	6.70	&				C2009	\\
2.521\(-\)0.220	&	17 52 21.17	&	 -26 53 21.1	&	-7.5	&	5.0	&	-6.1	&	1.02	&	4.2	&	0.70	&			2006dec04	\\
2.536+0.198	&	17 50 46.47	&	 -26 39 45.3	&	2.0	&	20.5	&	3.1	&	29.40	&	3.2	&	36.41	&				C2009	\\
2.591\(-\)0.029	&	17 51 46.69	&	 -26 43 51.2	&	-9.5	&	-4.0	&	-8.3	&	1.76	&	-8.2	&	1.69	&			2006dec04	\\
2.615+0.134	&	17 51 12.30	&	 -26 37 37.2	&	93.5	&	104.0	&	94.1	&	1.22	&	94.5	&	1.10	&			2006dec04	\\
2.703+0.040	&	17 51 45.98	&	 -26 35 56.7	&	91.5	&	98.0	&	93.5	&	8.97	&	93.6	&	9.00	&			2006dec04	\\
3.253+0.018	&	17 53 05.96	&	 -26 08 13.0	&	-1.5	&	3.5	&	2.2	&	3.54	&	2.2	&	3.70	&			2006dec04	\\
3.312\(-\)0.399	&	17 54 50.11	&	 -26 17 51.5	&	0.0	&	10.0	&	0.4	&	1.17	&	0.5	&	0.88	&			2006dec04	\\
3.442\(-\)0.348	&	17 54 56.11	&	 -26 09 35.6	&	-35.5	&	-34.5	&	-35.1	&	1.06	&	-35.0	&	0.66	&			2007jul19	\\
3.502\(-\)0.200	&	17 54 30.06	&	 -26 01 59.4	&	43.0	&	45.5	&	43.9	&	1.57	&	43.9	&	2.02	&			2006dec04	\\
3.910+0.001	&	17 54 38.75	&	 -25 34 44.8	&	15.0	&	24.5	&	17.8	&	5.04	&	17.9	&	5.10	&				C2009	\\
4.393+0.079	&	17 55 25.77	&	 -25 07 23.6	&	0.0	&	9.0	&	1.9	&	6.74	&	2.0	&	7.67	&			2006dec04	\\
4.434+0.129	&	17 55 19.74	&	 -25 03 44.8	&	-1.5	&	8.0	&	-1.0	&	3.29	&	-0.9	&	4.59	&			2006dec04	\\
4.569\(-\)0.079	&	17 56 25.30	&	 -25 03 03.7	&	9.0	&	10.0	&	9.5	&	0.44	&	9.5	&	0.61	&			2006dec04	\\
4.586+0.028	&	17 56 03.23	&	 -24 58 55.9	&	15.0	&	27.0	&	26.1	&	1.16	&	26.3	&	0.96	&			2006dec04	\\
4.676+0.276	&	17 55 18.34	&	 -24 46 45.3	&	-5.5	&	6.0	&	4.5	&	2.06	&	4.4	&	2.48	&			2006dec04	\\
4.866\(-\)0.171	&	17 57 25.96	&	 -24 50 24.4	&	5.0	&	6.0	&	5.4	&	0.56	&	5.4	&	0.64	&			2008aug19; (2008jan23)	\\
5.618\(-\)0.082	&	17 58 44.78	&	 -24 08 40.1	&	-28.0	&	-18.5	&	-27.1	&	3.37	&	-27.0	&	3.42	&			2006dec04	\\
5.630\(-\)0.294	&	17 59 34.60	&	 -24 14 23.7	&	9.0	&	22.0	&	10.5	&	1.28	&	10.6	&	1.31	&			2006dec04	\\
5.657+0.416	&	17 56 56.53	&	 -23 51 42.0	&	13.0	&	22.0	&	20.0	&	1.75	&	20.1	&	1.90	&			2006dec04	\\
5.677\(-\)0.027	&	17 58 39.98	&	 -24 03 57.2	&	-14.5	&	-11.0	&	-11.7	&	0.79	&	-11.5	&	0.99	&			2006dec04	\\
5.885\(-\)0.393	&	18 00 30.65	&	 -24 04 03.4	&	6.0	&	7.5	&	6.7	&	0.48	&	6.7	&	1.30	&			2006dec04	\\
5.900\(-\)0.430	&	18 00 40.86	&	 -24 04 20.8	&	0.0	&	10.6	&	10.4	&	6.20	&	10.4	&	6.20	&				C2009	\\

\end{tabular}
\end{table*}

The high spatial resolution of the ATCA (beam width of a few arcsec) 
has allowed us to recognise multiple sites that are blended together in 
the beam of the Parkes telescope used for the basic survey.  As discussed 
by Caswell (2009), the typical individual maser site with its own exciting 
star can have 
maser spots spread over an extent of up to 2 arcsec;  when maser spots are 
separated somewhat more than this, it is difficult to distinguish between 
a single site which is merely slightly more extended than usual, and a 
very small cluster of distinct sites, each with its own exciting star.
Where sites close to each other have been previously studied (e.g. Caswell 
2009) and been regarded as separate, we retain that interpretation. We 
have applied similar criteria to the new masers:  maser spots arising 
from a region up to about 2 arcsec in size are regarded as a single 
site;  notes to the sources 
indicate when the distinction is uncertain.  Where the Parkes spectrum   
is a blend of sites, the appropriate velocity range for each site has been 
determined from the ATCA observations.

In addition to the sources detected in the main survey, we also list for 
completeness a further five masers in this longitude range that are 
reliably documented but were not detectable in our survey;  for three of 
them we have made new measurements of the spectra, homogeneous with 
those of the 
other sources.  One of these sources (358.263-2.061) was absent from 
the initial survey because it lies just outside the survey latitude range; 
two others (346.517+0.117 and 348.579-0.920)  were below the survey 
noise limit but could be recognised in the longer integration MX follow up 
observations.  The remaining two sources (0.667-0.034 and 0.673-0.029, in 
the Sgr 2 complex) are not only weak but also very confused and we rely on 
the published report (Houghton \& Whiteoak 1995) based on an unusually 
deep survey.  As a result of these additions, in the longitude range 
presented here, we believe that there are no previously reported masers 
that are not accounted for by the entries of Table 1.

Only a few key parameters can be summarized in a table, and therefore we 
also present for each maser site a spectrum to display the detailed 
distribution of flux density as a function of velocity.  These spectra, 
shown in Fig. 1, are from Parkes observations, with half-power 
beamwidth (HPBW) 3.2 armin. At this spatial resolution, most sites are 
sufficiently isolated that their spectra are not confused.  However, there 
are other sites with much nearer neighbours, and their spectra are 
blended.  In these cases, the spectra are intended primarily to show the 
total emission from the combined region.  Reference to Table 1 is then  
required to recognise the strongest peak of each contributing site, and 
its velocity range.  Consultation of high spatial resolution data is 
needed to fully understand the most complex regions where neighbours have 
overlapping velocity ranges.  For some regions, such  
studies are already available (e.g. Houghton \& Whiteoak 1995 and Caswell 
1997);  other regions will be the 
subject of subsequent papers on the MMB survey (see Section 4.7), and 
preliminary assessment of complex regions is given in the remarks 
of the following section.    
We generally display the spectra from the lower noise `MX' follow-up 
observations, but in a few cases show the original survey observations 
instead, especially if the source was much stronger at the survey epoch.  
For the site 5.885-0.393, both survey cube and MX spectra are shown.

%Distinguish new sources and previously known sources? covered by c2009.

\subsection{Remarks on selected maser sites} 

Analysis of the ATCA data yields small maps of the maser spot spatial 
distribution at each site with relative accuracy of 0.1 
arcsec.  Presenting and interpreting the wealth of information in these 
maps is much beyond the scope of the present paper, but there are some 
sites where additional information from these maps has been vital in the 
generation of a reliable source catalogue, in particular, the 
interpretation of emission from closely spaced sites.   Accordingly, 
we have made remarks on selected maser sites in the following notes.    
A variety of other 
material also enhances the value of the catalogue and in this section we 
present individual notes for some of the sites.  In addition to a 
discussion of closely spaced sites, the remarks highlight the following:-   
the 5 masers not detectable in our original survey cubes; 
the 9 masers showing intensity variability greater than a factor of two 
within the survey and follow-up period;  16 other masers showing 
historical intensity variability; the 2 sources with widest velocity 
ranges, exceeding 16 \kms;  and the association of masers with an 
ultracompact \HII\ region (uc\HII) in some instances.    

Finally we remark on the likely location of some sites, especially
the 34 sources in this portion of the survey that lie in the 3-kpc arm  
(14 in the near side and  20 in the far side - see classification criteria 
in Section 4.6.2 and Green et al. (2009b)).  The suggested likely 
distances of other masers include some with respect to heliocentric 
distances, d, (distinguishing near and far kinematic distances);  
and others with respect to Galactocentric distances (R), chiefly those 
with longitude $\mid l \mid \le 3.4^{\circ}$, 
in order to highlight those that appear to lie within 3 kpc of the 
Galactic Centre, applying velocity criteria described in Section 4. 
\\

\subparagraph{345.003-0.223 and 345.003-0.224}  These known maser sites 
are shown on the same spectrum since they are separated by only 3 arcsec 
and 
do not overlap in velocity.  Reference to Table 1, and to spectra in 
Caswell (1997), make it clear which features 
are at which site.  Variability reported by Caswell et al. (1995b) is 
present at velocities corresponding to both sites, and spectra over a 
long timespan characterise the variability in more detail (Goedhart,
Gaylard \& van der Walt 2004) but reveal no distinct periodicities.  

\subparagraph{345.010+1.792 and 345.012+1.797}  These known maser sites 
are separated by 19 arcsec and are shown on the same spectrum.  The 
velocity 
ranges do not overlap (Table 1, and spectra from Caswell (1997)). The 
first site is associated with a uc\HII\ region (Caswell 1997) and displays 
class II methanol maser emission from a remarkably large number of other 
transitions (Cragg et al. 2001).

\subparagraph{345.198-0.030}   This new site is believed to lie in the far 
side of the 3-kpc arm.  For this site, and several others near these 
longitudes, the velocity (-1.5 \kms\ in this instance) is sufficiently 
near zero that the alternative kinematic distances lie on the solar 
circle, i.e. at the same Galactocentric distance as the Sun (and thus very 
nearby or at 17 kpc), where the space density of masers is low (see 
Section 4).  We conclude that, statistically, very few can lie near the 
solar circle, and that most of them are correctly attributed to the far 
side of the 3-kpc arm, but with some exceptions recognisable from their 
quite large Galactic latitude, e.g. 348.195+0.768.  

\subparagraph{345.407-0.952 and 345.424-0.951}  This is a known pair of 
sites  separated by about 1 arcmin.  The first is a single spectral 
feature, coincident with an OH maser site (Caswell 1998).  The spectra are 
shown aligned, one beneath the other and with the same velocity scale, so 
as to allow clear distinction of the features attributable to 
each site.   The sites are not likely to be at a large distance in view of 
their rather large Galactic latitude.  HI absorption measurements of 
nearby
continuum emission was interpreted to indicate a distance of 2 kpc by 
Caswell et al. (1975).  However, Radhakrishnan et al. (1972) considered 
this region in the context of others at similar latitude which lie  
between longitude  $345^{\circ}$ and $352^{\circ}$.  They argued 
persuasively that all of these complexes are at a distance of about 4.2 
kpc (if the Galactic Centre is assumed to be at 8.4 kpc) and this remains 
a likely interpretation.  

\subparagraph{345.441+0.205}  New site believed to lie in the far 
side of the 3-kpc arm.

\subparagraph{345.487+0.314 and 345.505+0.348}  Known sites separated by 
more than 2 arcmin.  The latter site displays emission over a velocity 
range from -23.1 to -10.5 \kms, and seems likely to lie in the far side of 
the 3-kpc arm;  with a peak intensity of 307 Jy, it is the strongest 
maser attributed to the far side.  The first site is a weak single 
feature with velocity -22.8 \kms\ lying just outside the velocity range of 
the 3-kpc arm (Green et al. 2009b).  For consistency, we formally exclude 
it 
from the 3-kpc arm but this is uncertain, and it may be near its apparent 
neighbour in the 3-kpc arm.  The spectra are aligned to allow recognition  
of the features of each site.

\subparagraph{345.498+1.467} This known site lies within the 
longitude-velocity domain of the far side of the 3-kpc arm but, in view of 
its large Galactic latitude, and evidence from HI absorption 
(Radhakrishnan et al. 1972), that interpretation is rejected in preference 
to a nearby location (Green 2009b).

\begin{figure*}
 \centering
\includegraphics[width=17cm]{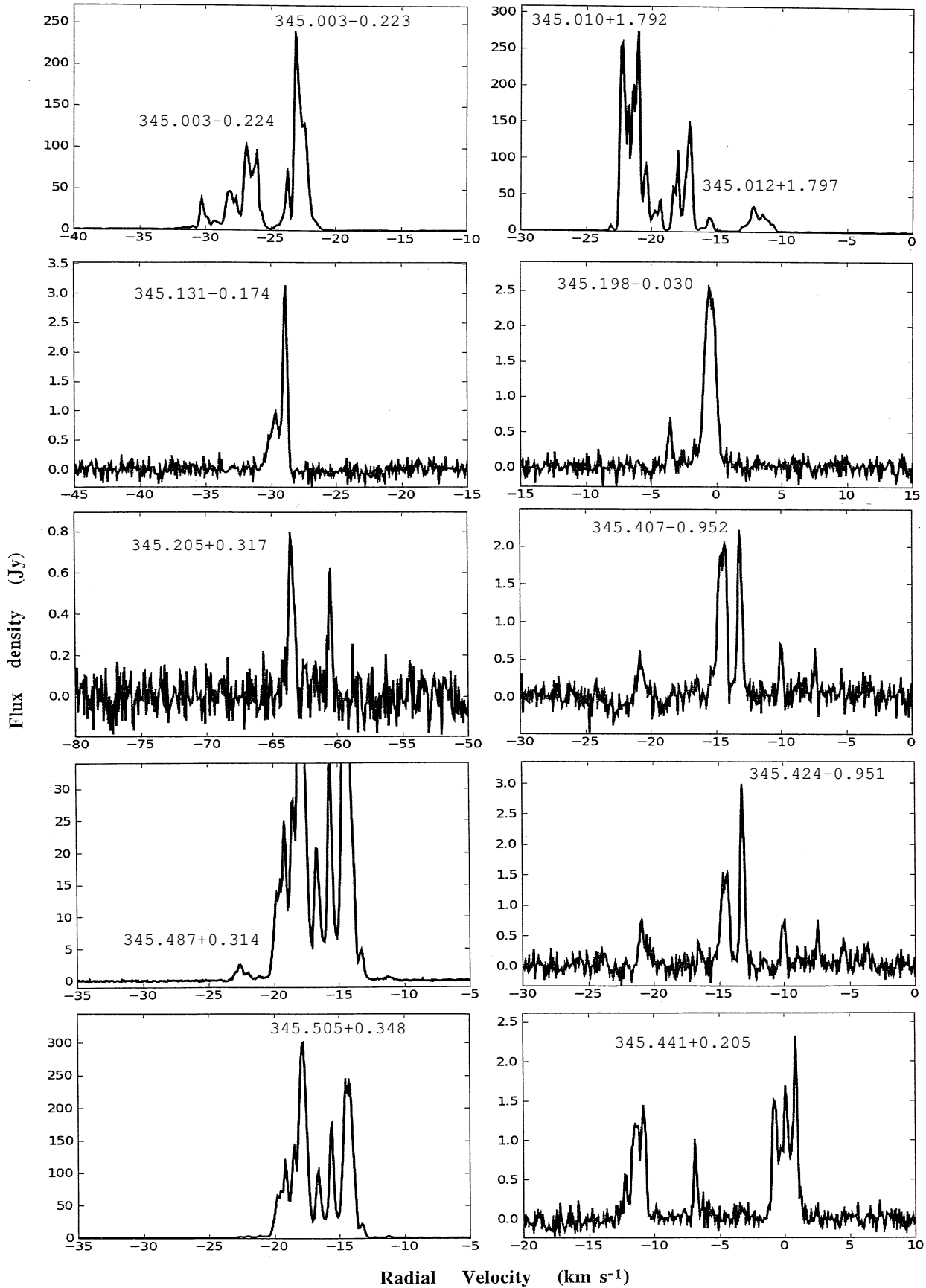}
\caption{Spectra of 6.6-GHz methanol masers. The combined emission from 
nearby sites is seen in some spectra and requires reference to Table 1 
and the text to resolve any confusion.  }

\label{fig1}

\end{figure*}

\begin{figure*}
 \centering

\addtocounter{figure}{-1}

\includegraphics[width=17cm]{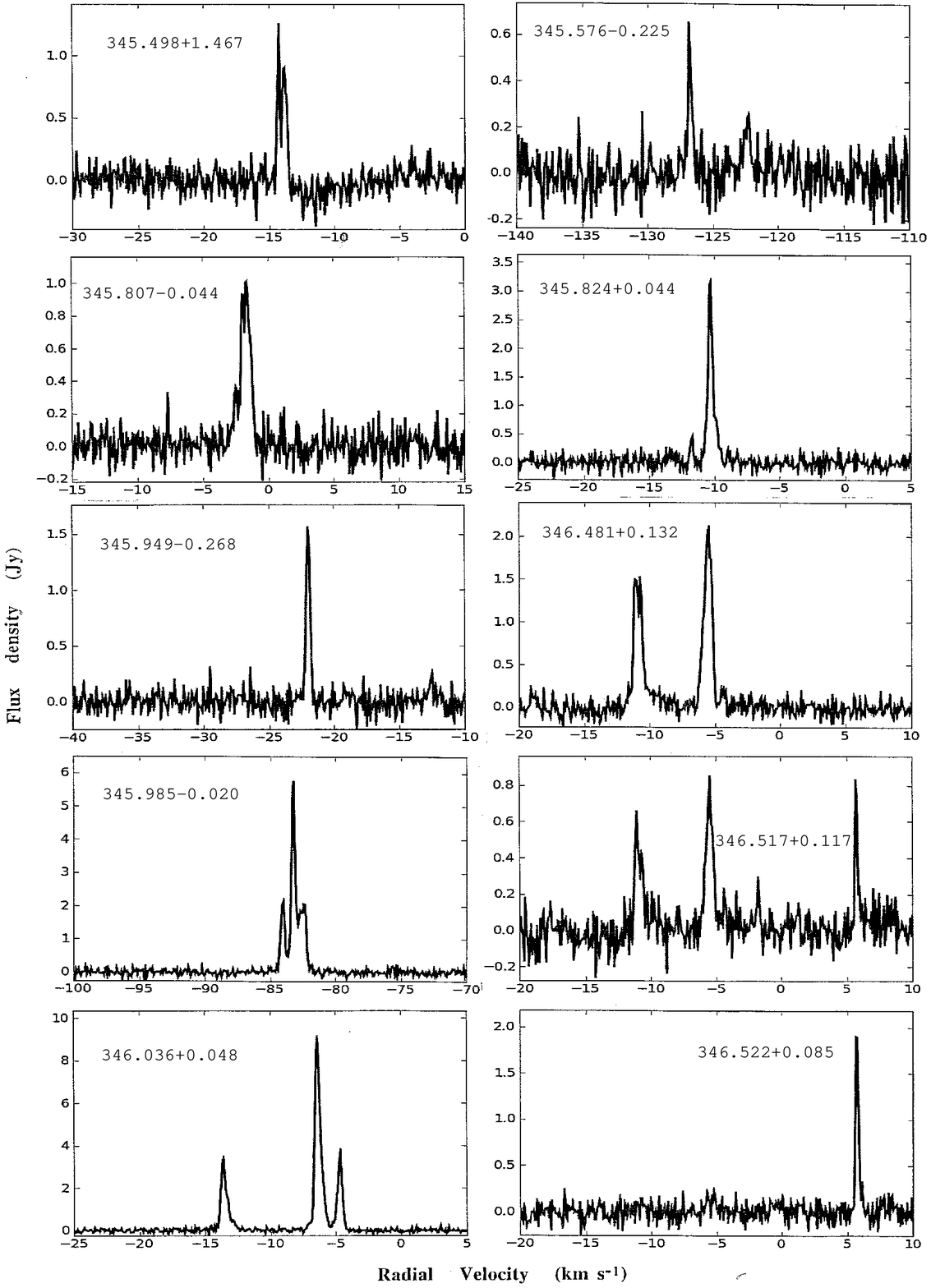}

\caption{\textit{- continued p2 of 16}}

%\caption{Spectra of 6.6-GHz methanol masers. }

\label{fig1} 

\end{figure*}

\begin{figure*}
 \centering

\addtocounter{figure}{-1} 

\includegraphics[width=17cm]{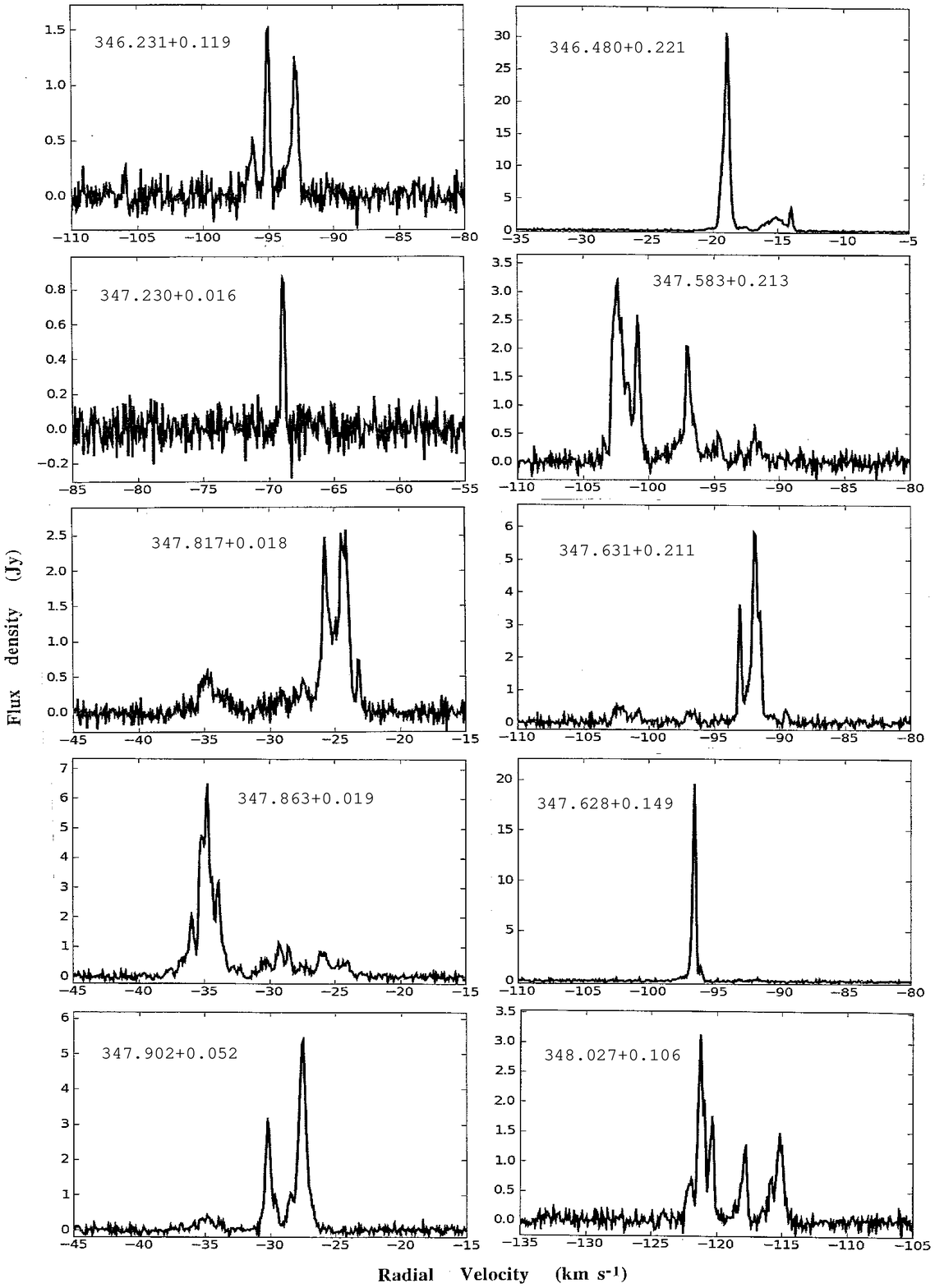}

\caption{\textit{- continued p3 of 16}}

%\caption{Spectra of 6.6-GHz methanol masers. }

\label{fig1}

\end{figure*}

\begin{figure*}
 \centering

\addtocounter{figure}{-1}  

\includegraphics[width=17cm]{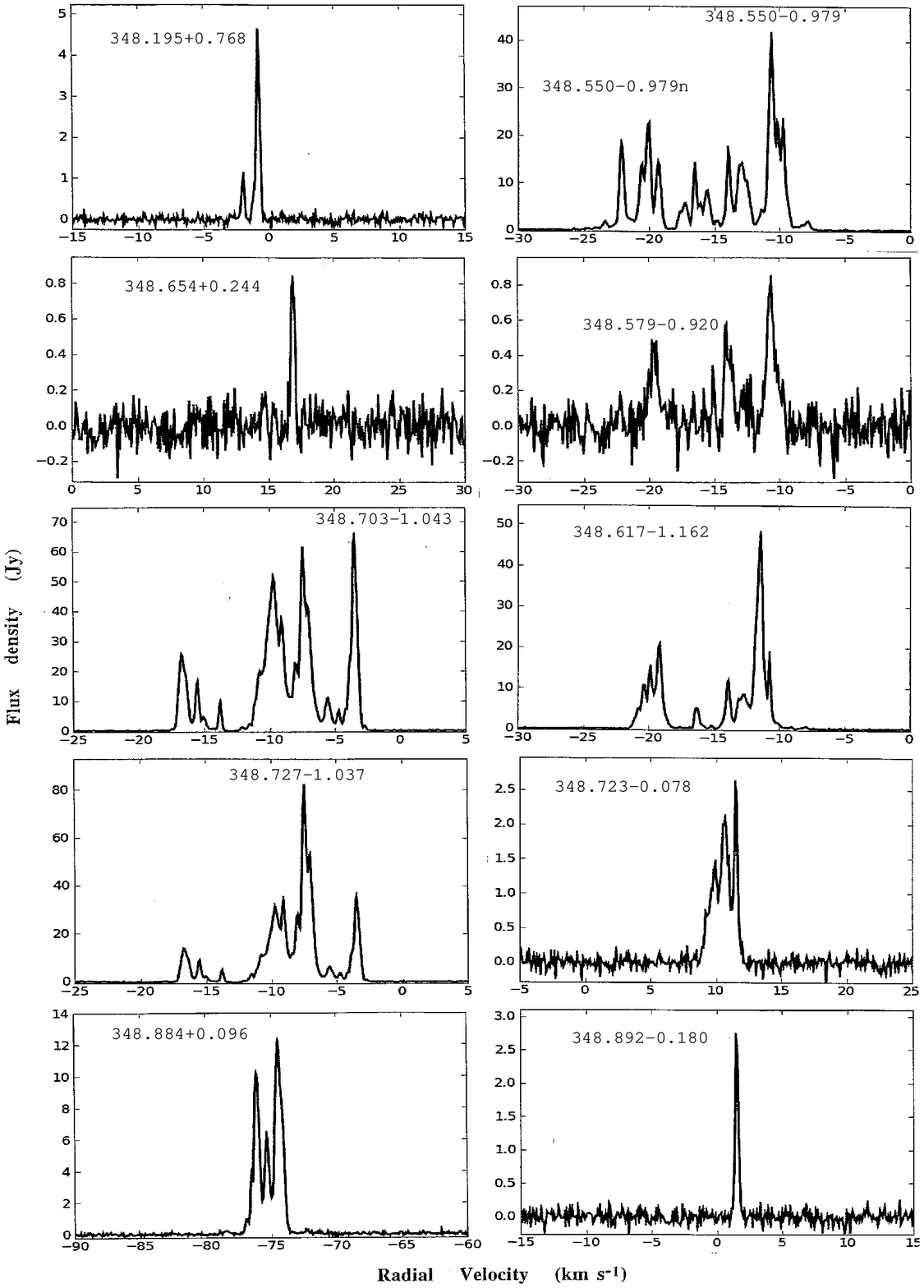}

\caption{\textit{- continued p4 of 16}}

%\caption{Spectra of 6.6-GHz methanol masers.  }

\label{fig1}

\end{figure*}

\begin{figure*}
 \centering

\addtocounter{figure}{-1}  

\includegraphics[width=17cm]{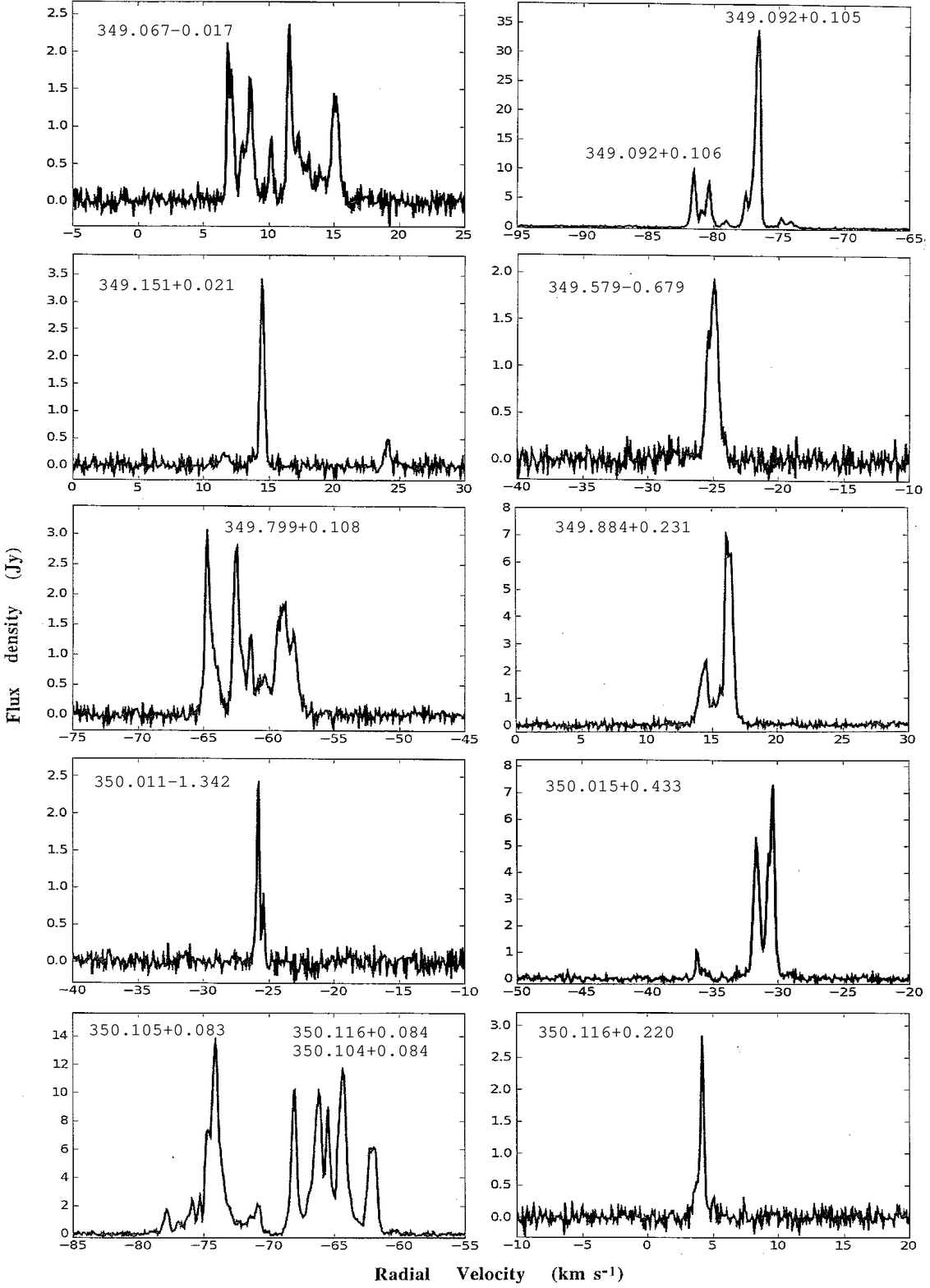}

\caption{\textit{- continued p5 of 16}}

%\caption{Spectra of 6.6-GHz methanol masers.  }

\label{fig1}

\end{figure*}

\begin{figure*}
 \centering

\addtocounter{figure}{-1} 

\includegraphics[width=17cm]{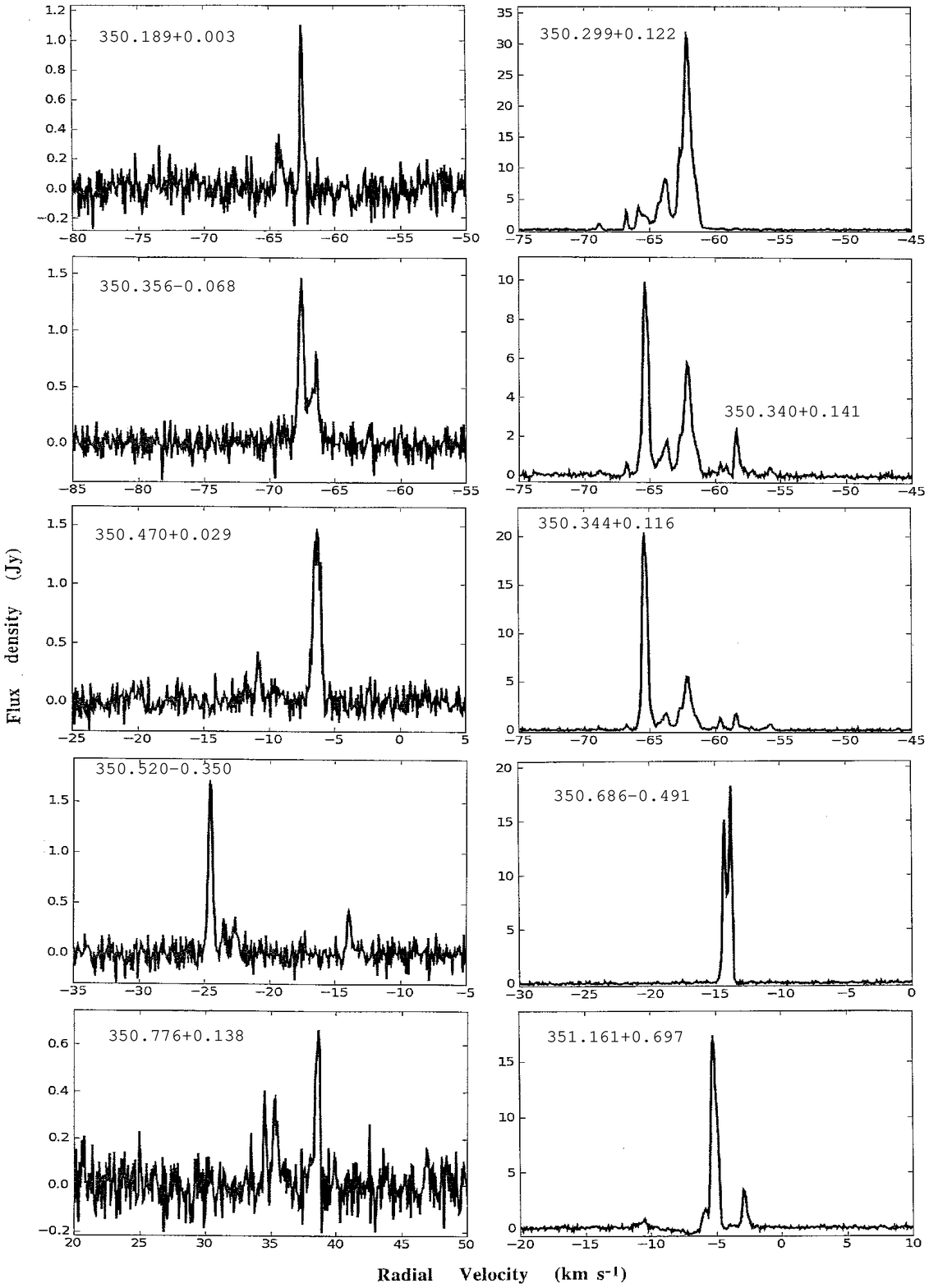}

\caption{\textit{- continued p6 of 16}}

%\caption{Spectra of 6.6-GHz methanol masers.  }

\label{fig1}

\end{figure*}

\begin{figure*}
 \centering

\addtocounter{figure}{-1} 

\includegraphics[width=17cm]{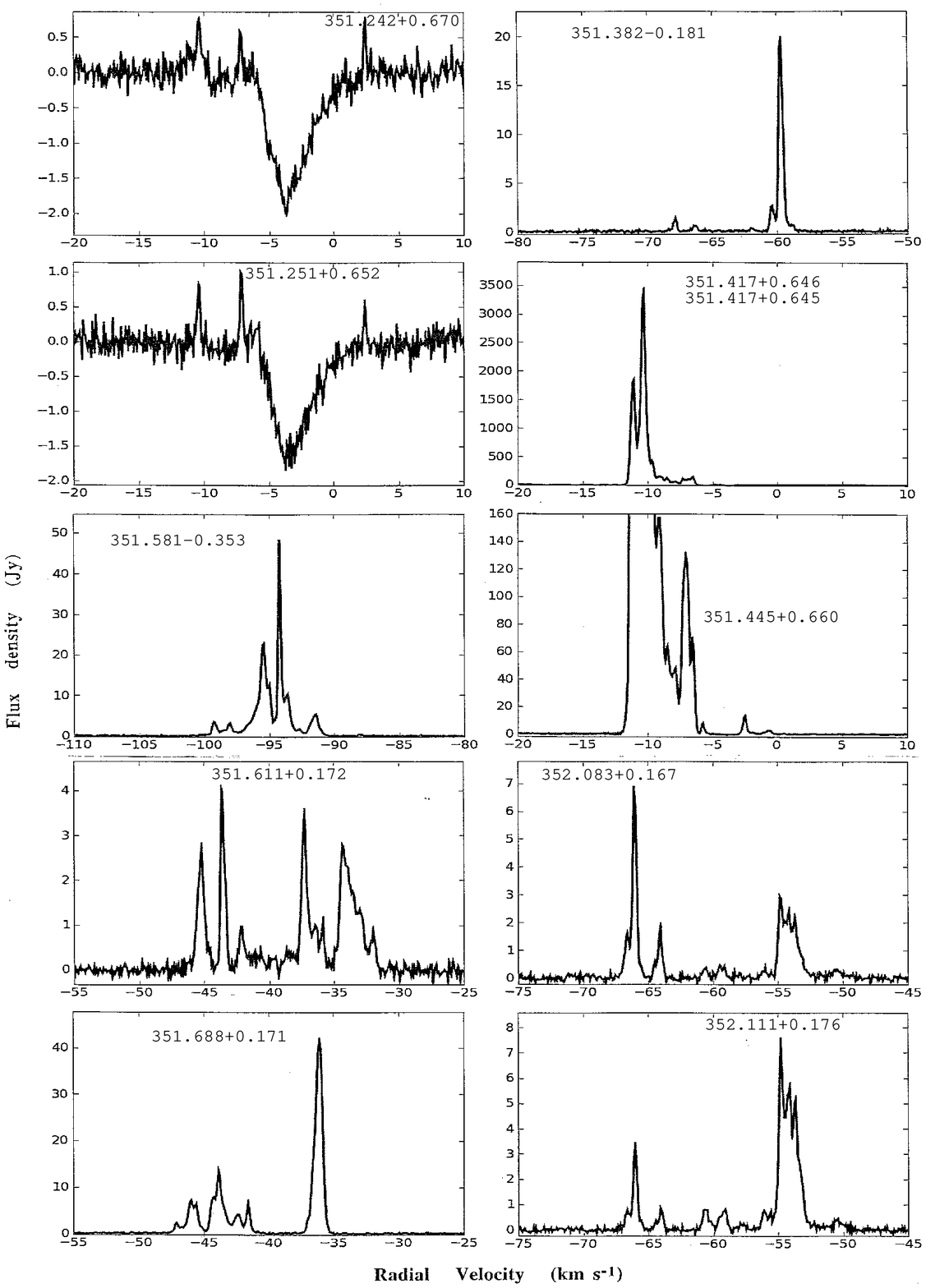}

%\caption{\textit{- continued p7 of 16}}

\caption{Spectra of 6.6-GHz methanol masers.  }

\label{fig1}

\end{figure*}

\begin{figure*}
 \centering

\addtocounter{figure}{-1} 

\includegraphics[width=17cm]{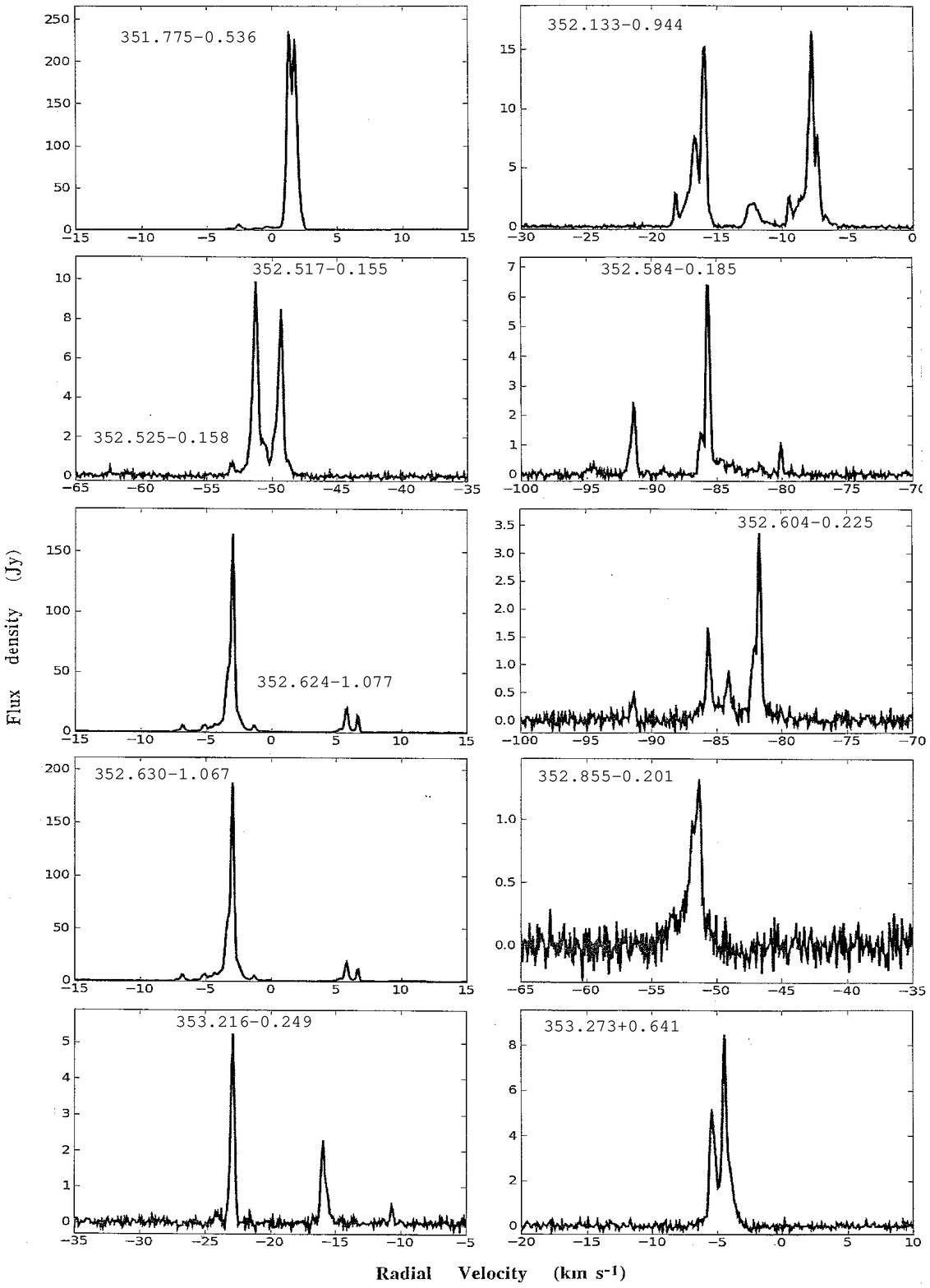}

\caption{\textit{- continued p8 of 16}}

%\caption{Spectra of 6.6-GHz methanol masers.  }

\label{fig1}

\end{figure*}

\begin{figure*}
 \centering

\addtocounter{figure}{-1} 

\includegraphics[width=17cm]{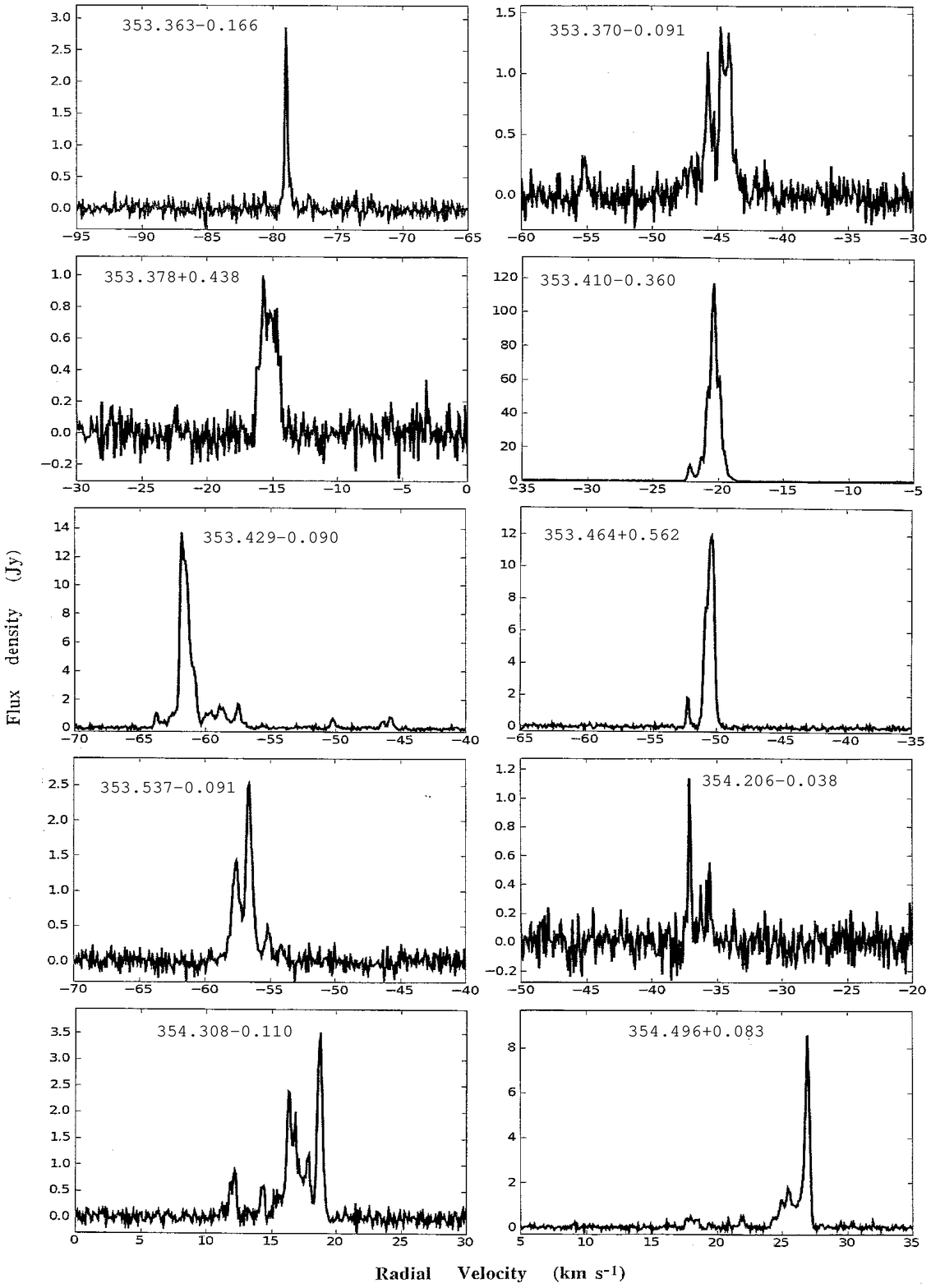}

\caption{\textit{- continued p9 of 16}}

%\caption{Spectra of 6.6-GHz methanol masers.  }

\label{fig1}

\end{figure*}

\begin{figure*}
 \centering

\addtocounter{figure}{-1}

\includegraphics[width=17cm]{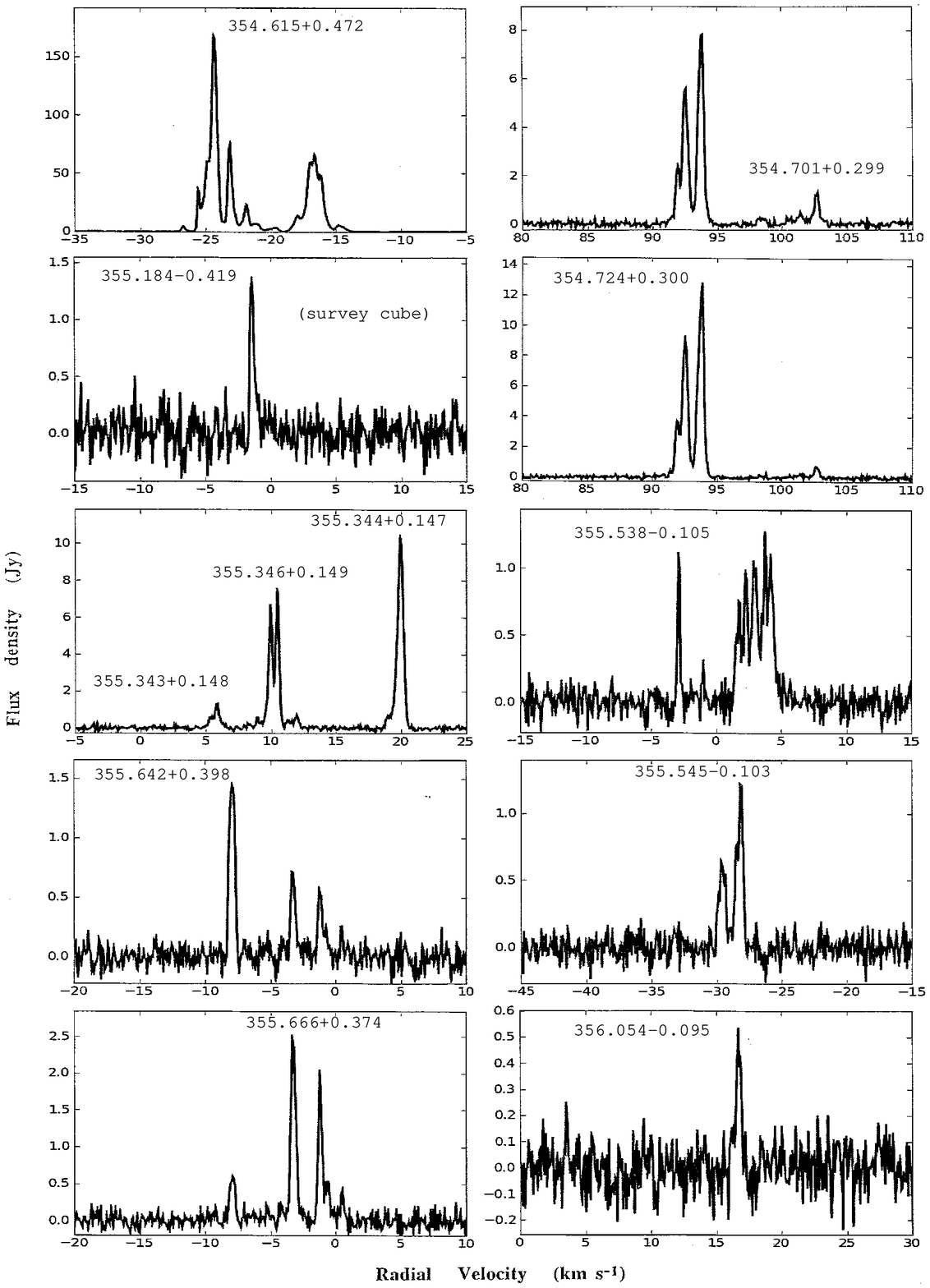}

 \caption{\textit{- continued p10 of 16}}

%\caption{Spectra of 6.6-GHz methanol masers.  }

\label{fig1}

\end{figure*}

\begin{figure*}
 \centering

\addtocounter{figure}{-1} 

\includegraphics[width=17cm]{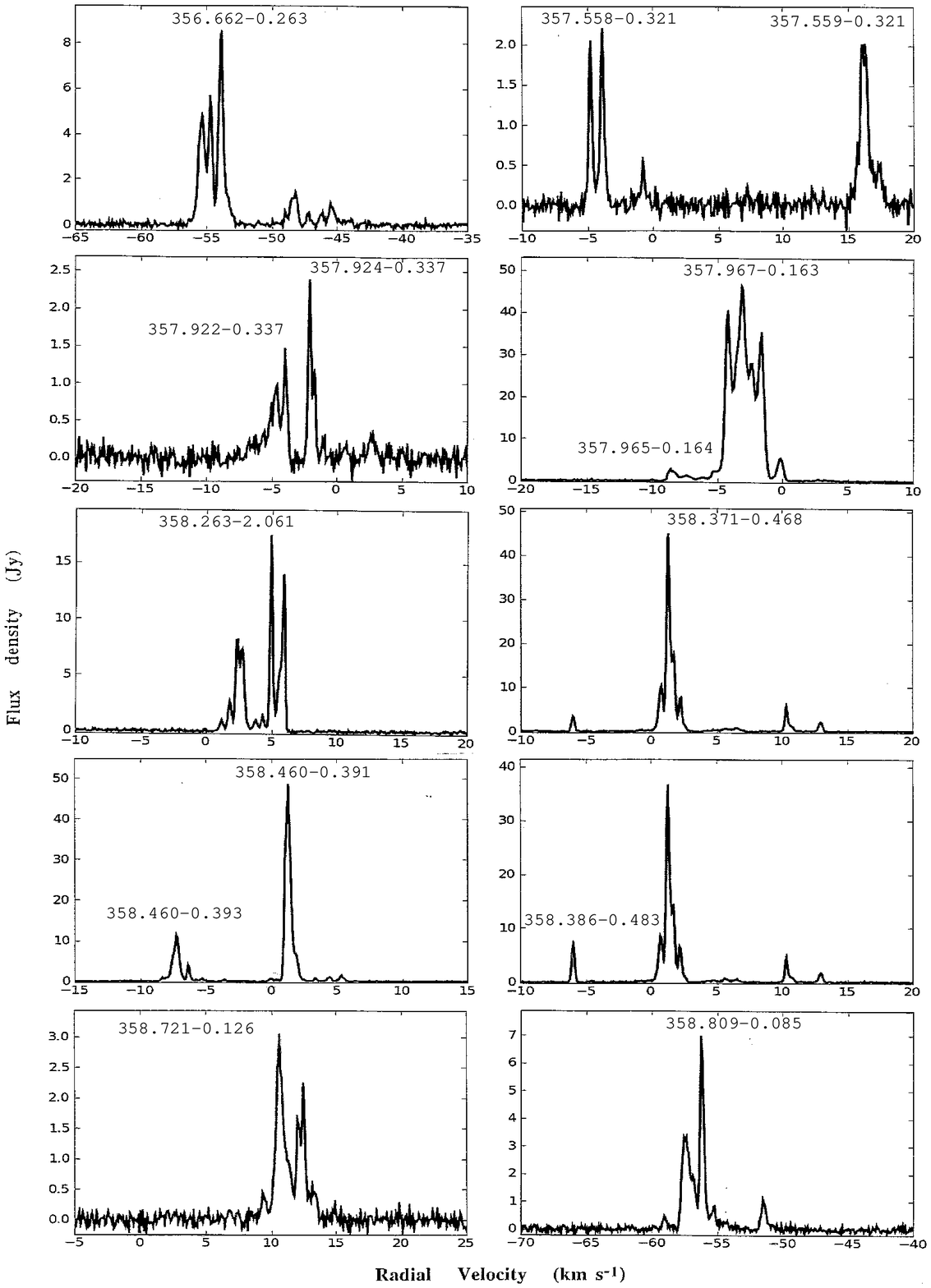}

 \caption{\textit{- continued p11 of 16}}

%\caption{Spectra of 6.6-GHz methanol masers.  }

\label{fig1}

\end{figure*}

\begin{figure*}
 \centering

\addtocounter{figure}{-1} 

\includegraphics[width=17cm]{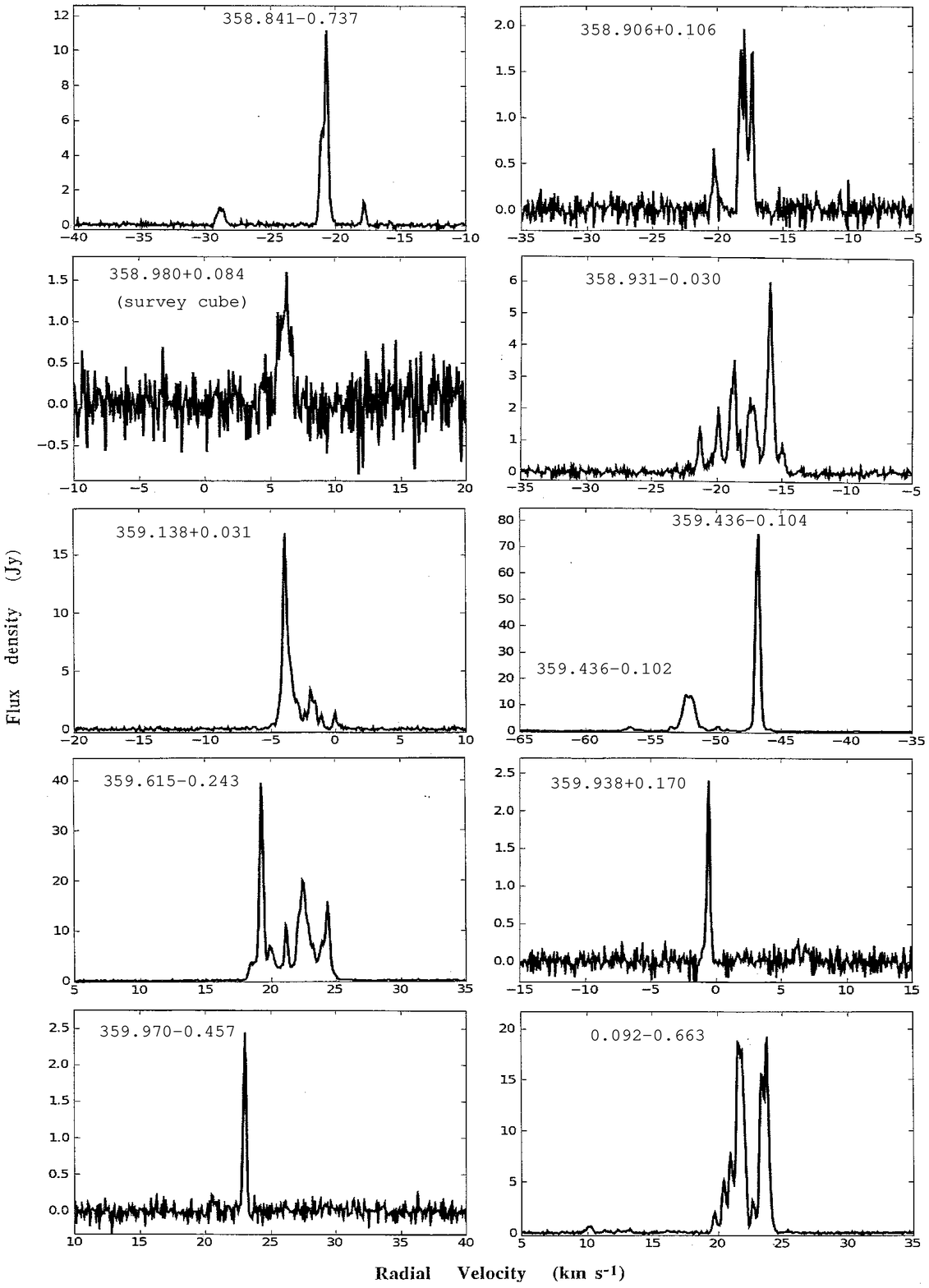}

\caption{\textit{- continued p12 of 16}}

%\caption{Spectra of 6.6-GHz methanol masers.  }

\label{fig1}

\end{figure*}

\begin{figure*}
 \centering

\addtocounter{figure}{-1} 

\includegraphics[width=17cm]{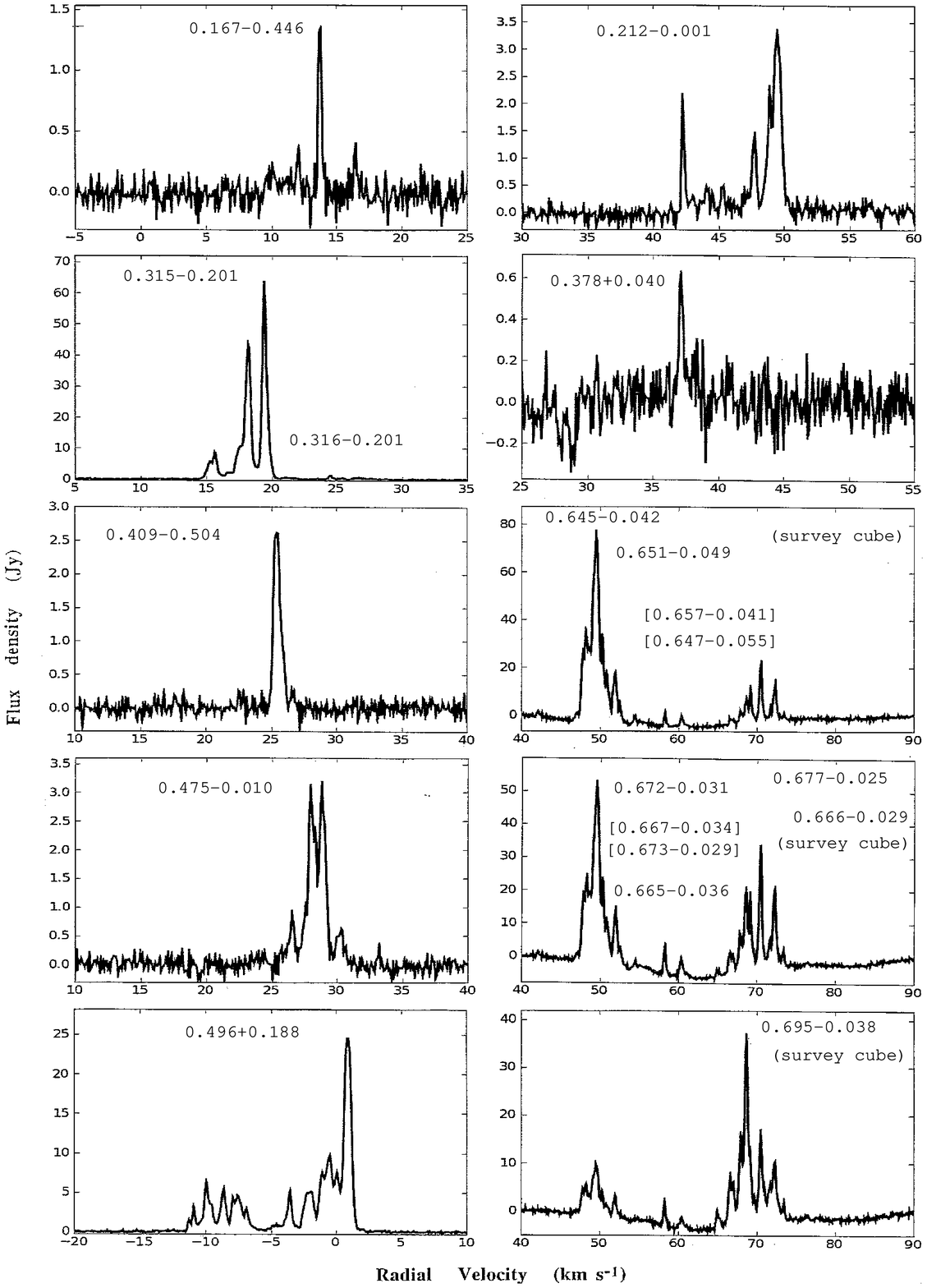}

 \caption{\textit{- continued p13 of 16}}

%\caption{Spectra of 6.6-GHz methanol masers.  }

\label{fig1}

\end{figure*}

\begin{figure*}
 \centering

\addtocounter{figure}{-1} 

\includegraphics[width=17cm]{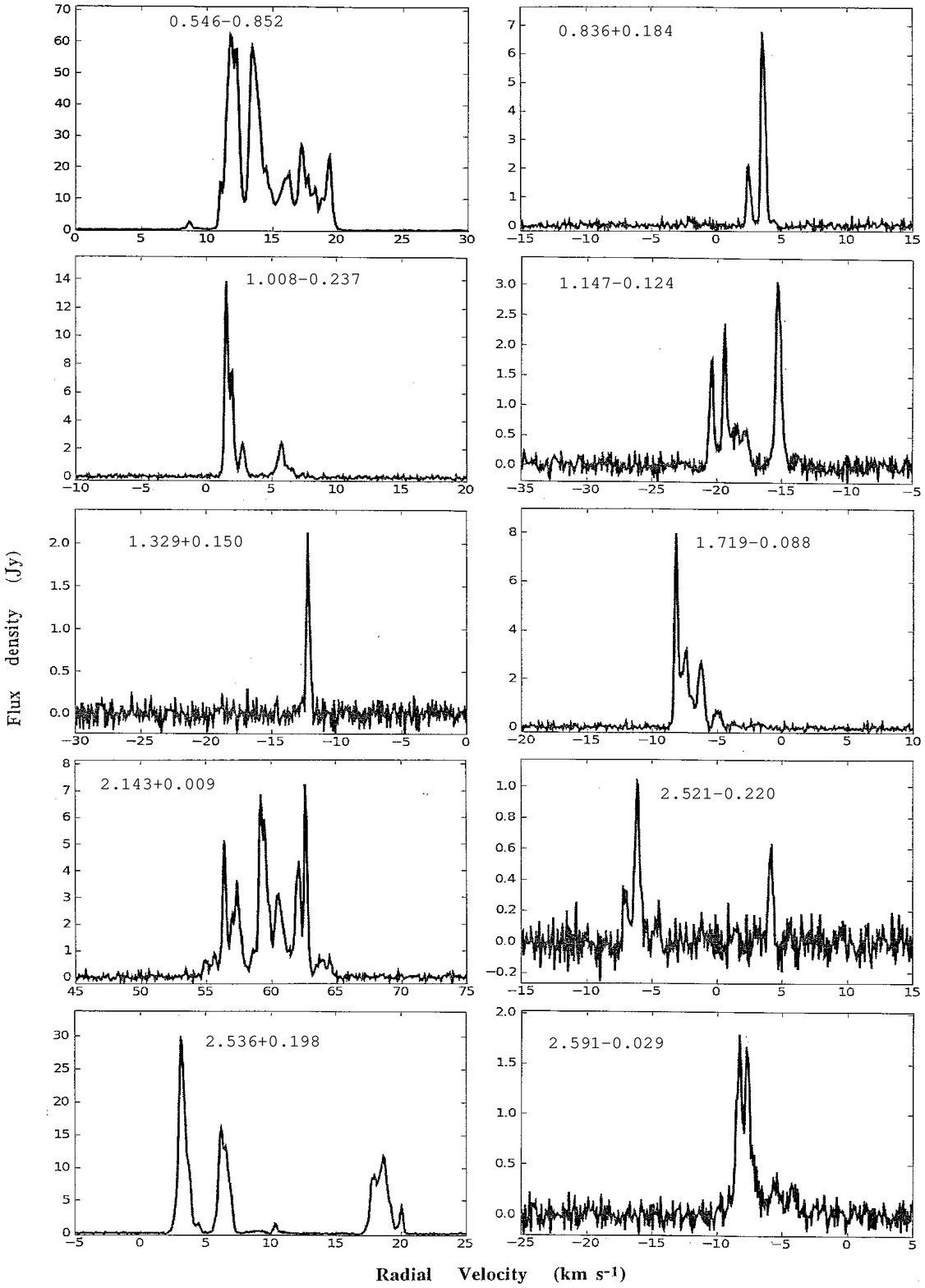}

\caption{\textit{- continued p14 of 16}}

%\caption{Spectra of 6.6-GHz methanol masers.  }

\label{fig1}

\end{figure*}

\begin{figure*}
 \centering

\addtocounter{figure}{-1} 

\includegraphics[width=17cm]{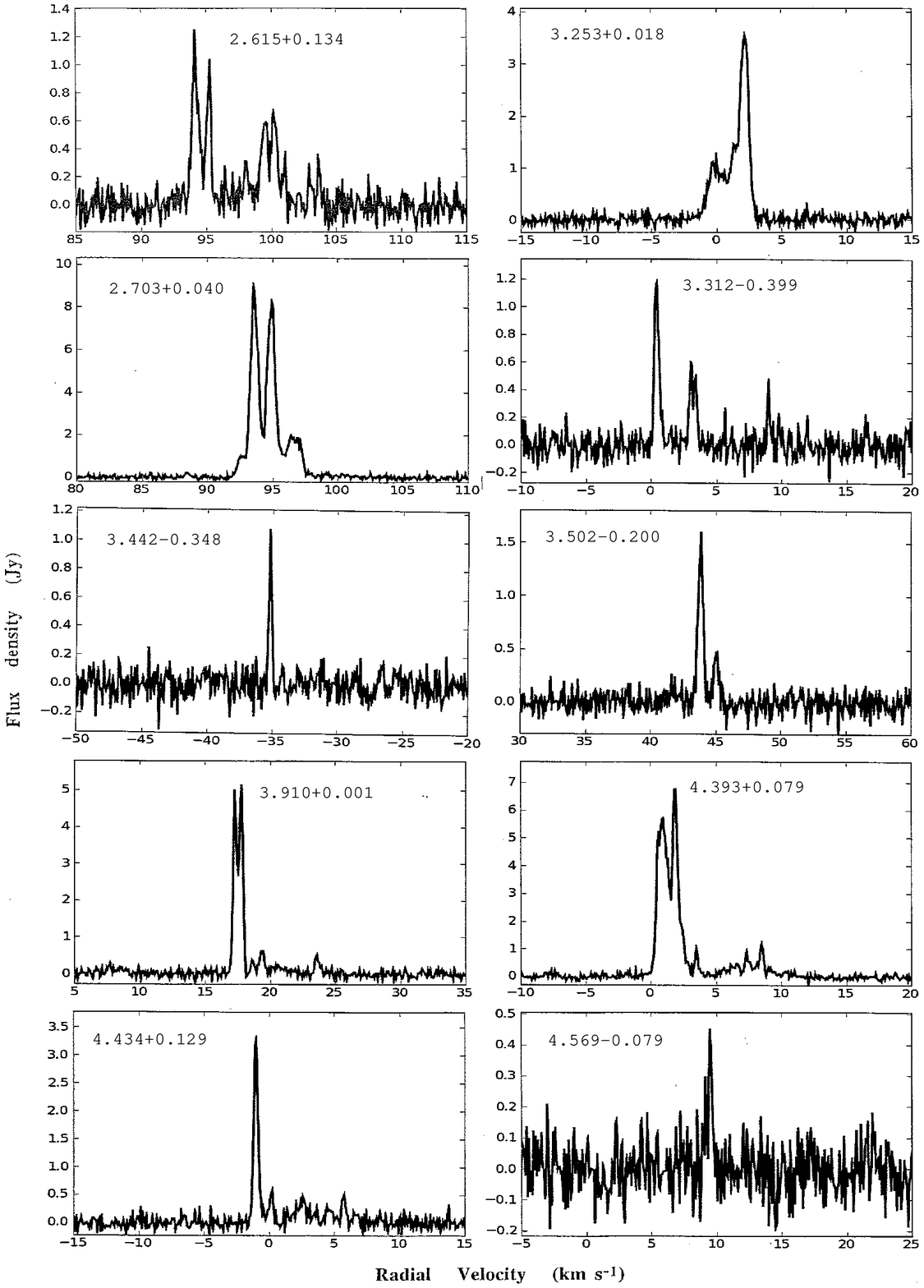}

\caption{\textit{- continued p15 of 16}}

%\caption{Spectra of 6.6-GHz methanol masers.  }

\label{fig1}

\end{figure*}

\begin{figure*}
 \centering

\addtocounter{figure}{-1}

\includegraphics[width=17cm]{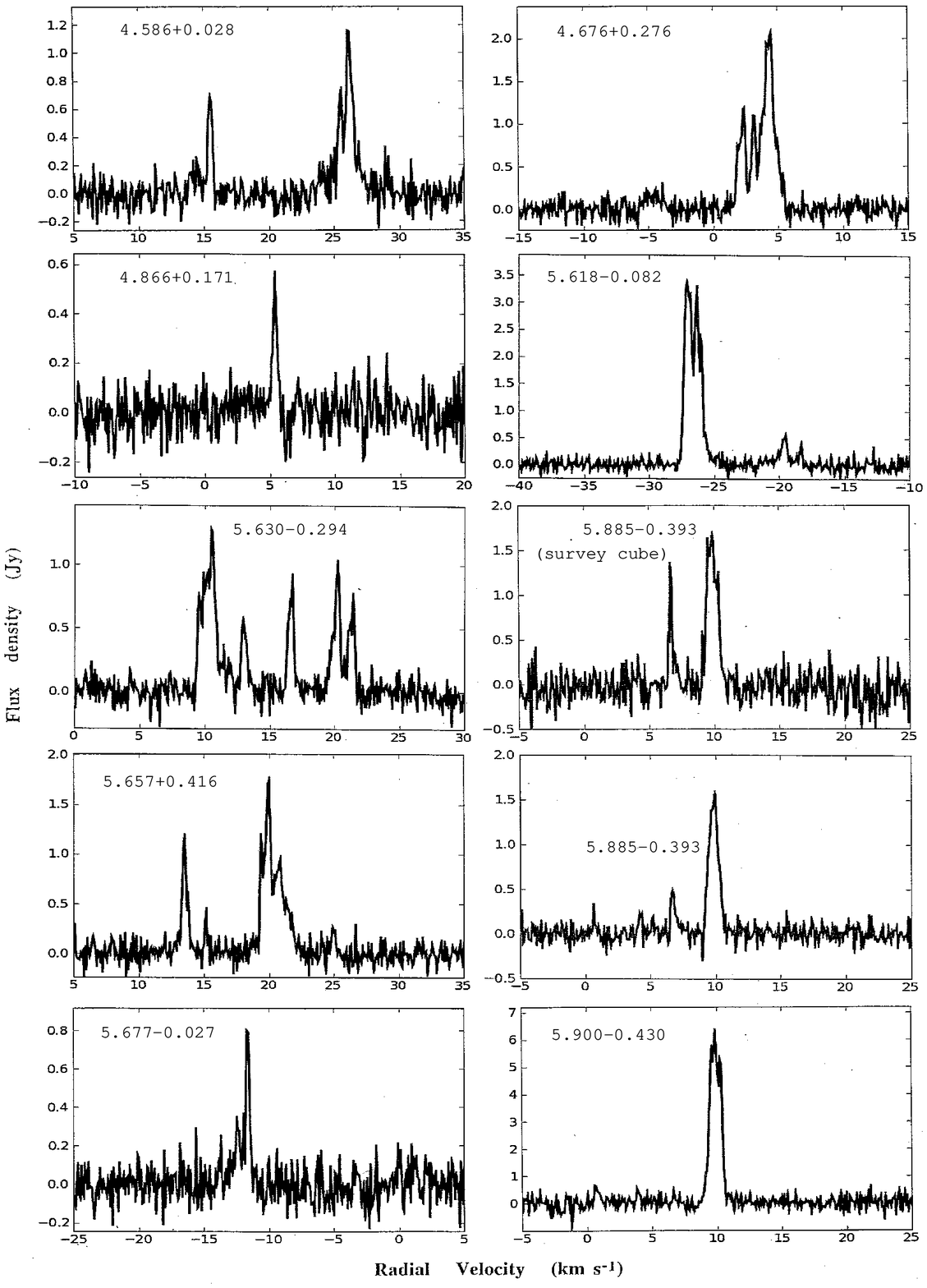}

 \caption{\textit{- continued p16 of 16}}

%\caption{Spectra of 6.6-GHz methanol masers.  }

\label{fig1}

\end{figure*}

\subparagraph{345.576-0.225}  This newly discovered site has the 
highest negative velocity (peak at nearly -126 \kms) of any site in the 
survey region presented here.  
It most likely lies in the near side of the expanding 3-kpc 
arm (Green et al. 2009b).  

\subparagraph{345.807-0.044}   New weak maser lying in the far side of 
3-kpc arm.  

\subparagraph{345.824+0.044}   Caswell and Haynes (1987) remark on the 
likely `far' distance of an \HII\ region which lies in the same 
direction and has the same velocity as the maser;  a location in 
the far side of the 3-kpc arm for both the \HII\ region and the maser 
would satisfy those observations.  

\subparagraph{345.985-0.020}  Strongly variable new maser with peak 
of 5.7 Jy at -83.2 \kms\ in our follow up measurements of 2008 August.  
This feature was only 1.35 Jy in our original survey cube (2007 
June, when another peak was slightly stronger) and had a similar value in 
the ATCA measurement of 2007 July.  

\subparagraph{346.036+0.048}  New maser located in the far side of the 
3-kpc arm.

\subparagraph{346.231+0.119}  This new maser has its velocity centred at 
-95 \kms, just outside our adopted range for 3-kpc arm near-side members;  
we provisionally reject it as a 3-kpc arm object pending other data.

\subparagraph{346.480+0.221}  Known site with velocity range just outside 
our formal boundary for far-side 3-kpc arm members.  

\subparagraph{346.481+0.132}  Known site located in the far side of 
the 3-kpc arm.  

\subparagraph{346.517+0.117}  Not detected above the noise level 
in the survey observations (2007 June) but known as a maser with peak of 1 
Jy at -0.1 \kms\ and a secondary peak of 0.7 Jy at -1.7 \kms\ in 1999 
October (see Caswell 2009).  Our follow-up observations 2008 August 
detected it with peak of only 0.3 Jy, and only at -1.7 \kms.  The spectrum 
is shown aligned in velocity with those of nearby sites 346.481+0.132 and 
346.522+0.085 so as to recognise their 
confusing contribution to the spectrum.  A similar confirmatory spectrum 
was obtained 2009 March.  

\subparagraph{346.522+0.085}   Known single feature varying in our 
observations between 1.47 Jy (2007 June) and 1.9 Jy (2008 August), similar 
to its peak in 1992, but as weak as 0.6 Jy in 1999 October (C2009).  
Located in the far side of the 3-kpc arm.  

\subparagraph{347.583+0.213}  Known maser in the near side of the 3-kpc 
arm.  The spectrum is aligned with 347.631+0.211 which is responsible for 
two weak features, and with 347.628+0.149 whose major feature does not 
contribute to the spectrum.   

\subparagraph{347.628+0.149}  Known maser in the near side of the 3-kpc 
arm.  The offset of 4.7 arcmin from the previous source is just large 
enough to prevent any confusion.  

\subparagraph{347.631+0.211}  Offset  3.7 arcmin from a possible 
companion, the previous maser 347.628+0.149, which is at slightly more 
negative velocity.  Most likely, 347.631+0.211 also lies in the near side 
of the 3-kpc arm, but at present, for consistency, we have formally 
rejected it as just outside our accepted velocity range.  Note that our 
spectrum centred here shows weak emission from 347.583+0.213 which lies at 
the edge of the Parkes telescope beam response.  

\subparagraph{347.863+0.019}  Our new observations of this site, both  
from the survey cube (2007 June) and the MX measurement (2008 August), 
show that, compared to the spectrum from 1992 June (Caswell et al. 1995a), 
the currently prominent features near -35 \kms\ have increased by a factor 
of 2 and the previously strongest feature of 7 
Jy at -29 \kms\ has faded to less than 1 Jy.   
The kinematic distance ambiguity for this site has been investigated by 
Busfield et al. (2006) who favour the far distance since no HI 
self-absorption could be detected.  

\subparagraph{348.027+0.106}  New maser located in the near side of 
the 3-kpc arm.  

\subparagraph{348.195+0.768}  New maser whose velocity is consistent 
with a location in the far side of the 3-kpc arm.  However its large 
latitude suggests that it is more likely to be nearby (Green et al. 
2009b).  If it is nearby, this should be easy to verify by 
future astrometry and a parallax measurement.

\subparagraph{348.550-0.979 and 348.550-0.979n} This is a known close pair 
of sources (separation 2 arcsec) for which we accept the current 
interpretation that they are 
distinct sites (Caswell 2009).  These sites, and those discussed in the 
next two notes, all have velocities that suggest they lie in a region 
argued to be at a distance of about 4.2 kpc by Radhakrishnan et al. (1972) 
- see also notes to 345.407-0.952.   

\subparagraph{348.579-0.920} This known weak maser displayed a peak of 0.5 
Jy at -15 \kms\ in 1996 (Caswell 2009) and was below the noise 
level in our regular survey measurement.  A weak peak of 0.32 Jy at this 
velocity was detected in 
our later follow-ups in both 2008 August and 2009 March and we 
conservatively list just this peak in Table 1.  However, the spectrum at 
this location, when aligned with that of the nearby strong source 
348.550-0.979, reveals that the sidelobe response to the strong source is 
no more than 1 per cent;  thus probably 
most other emission at this location is also from 348.579-0.920, and  
attributable to previously unrecognised features of up to 0.5 Jy.   

\subparagraph{348.617-1.162}  This is the strongest of our new detections 
and lies in the region where its velocity suggests it to be at a distance 
of 4.2 kpc (Radhakrishnan et al. 1972).  

\subparagraph{348.654+0.244}   New maser located in the far side of the 
3-kpc arm. 

\subparagraph{348.703-1.043 and 348.727-1.037}  This is a known pair of 
sites  separated by more than 1 arcmin.  Both sites host OH masers 
(Caswell 1998).  Emission from the first site is confined to a small 
velocity range, and its main feature at velocity -7.4 \kms\ is clearly 
distinguishable from 348.727-1.037 when the two aligned spectra are 
compared.  The sites lie in the region suggested to be at a 
distance of 4.2 kpc (Radhakrishnan et al. 1972).  

\subparagraph{348.723-0.078, 348.892-0.180, and 349.067-0.017}   The first 
is a new maser and the other two are previously known.  All are most 
likely located in the far side of the 3-kpc arm.    

\subparagraph{349.092+0.105 and 349.092+0.106} Known sites with separation 
of only 2 arcsec.  Their velocities near -80 \kms\ are not quite 
within the range of the near side of the 3-kpc arm.  If they are not 
in the 3-kpc arm, their velocity would be compatible with a location 
slightly further away, and within 3 kpc of the Galactic 
Centre.  At a distance of 8.4 kpc, their separation corresponds to 
more than 80 mpc.  

\subparagraph{349.151+0.021}  New site located in the far side of the 
3-kpc arm.  

\subparagraph{349.579-0.679} This new maser has decreased from a peak 
flux density of 5.9 Jy in the initial survey (2007 June) to 4.2 Jy in the 
ATCA observations (2007 July) and 1.9 Jy (2008 August) in our MX spectrum.

\subparagraph{349.884+0.231}  New maser located in the far side of the 
3-kpc arm.

\subparagraph{350.011-1.342}  Known maser consistently showing a peak 
of more than 2 Jy in our recent observations.  However,  
the original detection of 1993 September (reported in source notes of 
Caswell 1998) showed a peak of 1 Jy at velocity 
-27.9 \kms, and in 1999 May the peak was only 0.4 Jy at -25.8 \kms\ 
(Caswell 2009);  thus it has shown a recent flux density increase by more 
than a 
factor of five.  Its large latitude suggests that it is nearby, but the 
velocity and latitude would be compatible with a location in the extensive 
structure at 4.2 kpc discussed by Radhakrishnan et al. (1972).  

\subparagraph{350.105+0.083, 350.104+0.084 and 350.116+0.084}  Three 
known  sites (Caswell 2009), of which the first two are separated by only 
4 arcsec.  The 
first site is the strongest, with a large velocity range, and a peak at 
-74.1 \kms\ of approximately 15 Jy, both in our recent measurements up to 
2008 March and as far back as 1996 October; however it was much stronger, 
40 Jy, in 1992.  The location of the second site was determined in 1996 
October from a single feature at -68.4 Jy of 2.5 Jy, and the location of 
the third site, offset 40 arcsec, was determined from a single 
feature in 1996 October at -68.0 \kms, with peak 1.8 Jy.  
A much stronger feature close to these velocities now has a peak varying 
between 14.6 (survey cube 2007 March) and 9.9 Jy (MX 2008 March), and 
although we have no precise recent position, we interpret it as evidence 
of a flare in either the second or third site by a factor of four.  Slight 
differences in the strength of emission seen in our Parkes measurements 
centred at each site are insufficient to indicate the more likely 
location.  However, from the velocity comparisons, it seems probable  
that the flare occurred in the third site rather than the second.   

\subparagraph{350.116+0.220}   New maser located in the far side of 
the 3-kpc arm.  

\subparagraph{350.340+0.141 and 350.344+0.116} 350.340+0.141 is a single 
feature weak new maser nearly 90 arcsec from the known strong maser 
350.344+0.116.  The aligned spectra clearly reveal the main features of 
each site.  

\subparagraph{350.776+0.138}  The high positive velocity of this new site 
is strikingly unusual for a site at this longitude, and can most readily 
be interpreted as indicating a location at the edge of the far side of the 
3-kpc arm. 

\subparagraph{351.242+0.670 and 351.251+0.652}  The first of these was 
reported by Caswell and Phillips (2008) while the second was detected in 
the same observations (but unpublished) as a peak of 1 Jy at -7.1 \kms.   
Neither maser can readily be seen in our Parkes survey due to a 
combination of several distracting effects.  These effects include deep 
absorption (stronger than the masers) at a velocity spanning almost the 
entire range between the two masers; the consequent slightly offset zero 
intensity level;  and some weak responses to very strong offset maser 
emission near velocity -10 \kms\ associated with NGC6334F.  
Allowing for these effects, our original survey cube for 351.242+0.670 
shows a peak of 2.2 Jy at +2.5 \kms.  The follow-up MX spectrum 2008 
March (displayed in this paper) shows a peak of 0.74 Jy.  At intermediate 
epochs, the peak was 1.1 Jy on 2007 May 14 using a compact configuration 
of the ATCA (Caswell \& Phillips 2008), and our high spatial resolution 
measurements with the ATCA on 2007 July 21 revealed a peak of only 0.44 
Jy.  The maser is thus markedly variable.  

The second site is offset 73 arcsec from the first and is weaker.  The 
most sensitive spectrum from the MX follow-up (2008 March) shows a 1-Jy 
peak at -7.1 \kms, with weaker emission extending to -6 \kms, similar to 
the unpublished ATCA observations by Caswell \& Phillips in 2007 May;  the 
original survey cube was noisier, showing a peak of 1.3 Jy, and the ATCA 
observations (2007 July) yielded a peak of 0.6 Jy at -7.1 \kms, with a 
secondary peak of 0.55 Jy at -6.1 \kms.   Some variability seems likely, 
but deviations from a peak 
of 1 Jy are barely significant in view of the low signal-to-noise ratio.  
The two sites do not overlap in their emission velocity ranges and are 
clearly distinguishable on the aligned spectra.

\subparagraph{351.417+0.645, 351.417+0.646 and 351.445+0.660} These are 
strong well-known masers (e.g. Caswell 1997) with the first projected 
onto the prominent compact \HII\ region NGC6334F (Ellingsen, Norris \& 
McCulloch 1996;  Caswell 1997).  The second is offset from the first 
by several arcsec to the north-west and has an almost identical velocity 
range but the spectra are clearly distinct when seen with the ATCA spatial 
resolution (Caswell 1997).  The third site is offset nearly 2 arcmin to 
the north, and our Parkes spectrum, although confused by the first 
two very strong sources between -12 and -9.6 \kms, clearly shows features 
at -9.2, -7.1 and -2.5 \kms\ which correspond to the main features of 
351.445+0.660 seen on the ATCA spectrum of Caswell (1997).

\subparagraph{351.581-0.353}   This known strong maser site has a 
weak northern feature offset nearly 2 arcsec (Caswell 1997; 2009).  
Following Caswell (2009) we treat it as probably not a distinct site and 
therefore do not list it here, and cite a velocity range encompassing 
all features.  The velocity indicates a location in the near side of the 
3-kpc arm.  

\subparagraph{351.611+0.172 and 351.688+0.171} These masers are separated 
by 4.6 arcmin and both are new.   351.688+0.171 has a peak flux density of 
42 Jy, our second strongest new detection.  
Alignment of the two spectra shows that, despite the similar velocity 
ranges, there is no confusion between them in the Parkes spectra.  Both 
sites are likely to be part of the same star forming cluster.  

\subparagraph{352.083+0.167 and 352.111+0.176} This known pair of sites 
has a separation of more than 100 arcsec and non-overlapping velocity 
ranges.  The intensity of the second site has been stable but the first 
has increased by more than a factor of 3 relative to 1999 May and 
1993 September (Caswell et al. 1995a; Caswell 2009).  The aligned spectra 
clearly distinguish the features from each site.  

\subparagraph{352.604-0.225 and 352.584-0.185}   Two new sites 
separated by more than 2 arcmin, with the velocity range of the stronger 
site, 352.584-0.185, straddling that of the weaker site.  Their systemic 
velocities correspond well with that expected for the near side of the 
3-kpc arm.  Both show some variability.   The aligned spectra 
show a small amount of confusion between them as measured at Parkes, with 
a beamsize of 3.2 arcmin. 

\subparagraph{352.624-1.077 and 352.630-1.067} Two known sites separated 
by 42 arcsec.  The velocity ranges do not overlap and the separation of 
the sources is clear on the aligned spectra.  

\subparagraph{353.216-0.249} New source with large variability, 
increasing from the survey cube peak of 1.1 Jy (2007 March) to 
1.3 Jy (ATCA, 2007 July) and reaching 5.1 Jy in the MX follow-up (2008 
March).

\subparagraph{353.273+0.641} Known source (Caswell \& Phillips 2008) 
associated with an unusual water maser showing dominant blue-shifted 
emission.  The methanol intensity has faded from 25 Jy in 1993 June to 
12.7 Jy in the survey cube (2007 March) and finally to 8.3 Jy in the MX 
follow-up 2008 March.  

\subparagraph{353.363-0.166}   New maser located in the near side of 
the 3-kpc arm.  

\subparagraph{353.429-0.090} New maser with large velocity range of 18.9 
\kms, the widest in this part of the survey.  However, the weak features 
in the velocity range -51 to -45 \kms\ were not detected above the noise 
level of the ATCA observations so it was not possible to confirm 
conclusively that they all arise from this same site.   

\subparagraph{354.496+0.083}  New maser located in the far side of the 
3-kpc arm.  

\subparagraph{354.615+0.472}  Known strong maser, showing modest 
variations in our survey and follow-up period.  However the feature at -23 
\kms, which had a peak intensity of 216 Jy in 1992 December, has now faded 
to less than 100 Jy, whereas some other features show negligible change.  
The high variability was noted by Caswell et al. (1995b) and additional 
monitoring information was presented by Goedhart et al. (2004).  

\subparagraph{354.701+0.299 and 354.724+0.300}  The first of these is a 
new maser with high positive velocity exceeding 100 \kms;  the second 
maser was already known and has a similar velocity.  Such velocities are 
highly unusual for this region of Galactic longitude.  Indeed, they are 
the highest positive velocity masers in any part of the survey region 
covered here.  The unusual velocities are discussed in Section 4.5, in the 
context of the Galactic bar which appears to be their most likely 
location.  The separation between the sites is 75 arcsec, causing spectral 
features of the offset source to be seen on both Parkes spectra.   

\subparagraph{355.184-0.419} This is a new maser for which we display the 
spectrum from the original survey cube (2007 January).  Confirmation 
observations with the ATCA (2007 February) showed a peak of 1.2 Jy and 
yielded a precise position.  No MX measurement has yet been made.  

\subparagraph{355.343+0.148, 355.344+0.147 and 355.346+0.149}  These
are three previously known sites, with the first two separated by only 
about
2 arcsec but the third offset by 10 arcsec.  Recent careful evaluation of
information that in the past gave rise to conflicting interpretations of
their distance has now yielded the conclusion that the masers most likely
lie within 3 kpc of the Galactic Centre, or perhaps the far side of the   
3-kpc arm (Caswell 2009).  A single spectrum is shown, and reference to   
the table, and to ATCA spectra from Caswell (1997), is needed to 
distinguish the features.

\subparagraph{355.642+0.398 and 355.666+0.374}  New sites, separated 
by just 
over 2 arcmin.  Their velocity ranges do not overlap and the aligned 
spectra distinguish the two sites.

\subparagraph{356.662-0.263}    Known maser site located in the 
near side of the 3-kpc arm.

\subparagraph{357.558-0.321 and 357.559-0.321} Newly discovered pair of 
sites separated by 3 arcsec and with quite separate velocity ranges.  The 
second source has velocity range +15 to +18 \kms\ and is best interpreted 
as being at Galactocentric radius R $<$ 3 kpc (see Section 4.6).  
The first source has a velocity range of -5.5 to 0 \kms; from this 
velocity, its location could be interpreted as R $>$ 3.5 kpc.  However 
their small projected separation suggests that the sites are more likely 
close companions, at a common distance.  Therefore, for both sites, we 
favour  R $<$ 3 kpc, where anomalous velocities are more common than at R 
$>$ 3.5 kpc.  

\subparagraph{357.922-0.337 and 357.924-0.337} Another newly discovered 
pair, separated by 5 arcsec, with slight overlap of velocity ranges.
The velocities are compatible with a location  R $>$ 3.5 kpc.  

\subparagraph{357.965-0.164 and 357.967-0.163} This known pair of sites 
has a separation of 7 arcsec.   Intensity variations have been modest for 
all features. The stronger emission of the second source lies wholly 
within the velocity range of the (much wider velocity) weak source.  Their 
velocities are compatible with a location  R $>$ 3.5 kpc.  

\subparagraph{358.263-2.061.} The Galactic latitude of this previously 
known site is large, and lies just outside our formal survey coverage.  We 
have re-observed it and include it in Table 1 so that the table provides 
a complete current listing of reliably known masers in the survey 
longitude range.  The maser intensity has shown modest variations. Its 
velocity, ranging from 0.5 to 6 \kms, is compatible with R $>$ 3.5 kpc, 
after allowance for small (less than 7 \kms) non-circular motions.  This 
is consistent with its large latitude, which almost certainly indicates 
that it is nearby, with heliocentric distance of only a few kpc.  

\subparagraph{358.386-0.483 and 358.371-0.468} Both sites are previously 
known, separated by about 81 arcsec.  The first shows emission from a 
single narrow peak at velocity  -6.2 \kms\ which was only 2.5 Jy in 1997 
May (Caswell 2009) but the survey cube peak was 12.5 Jy in 2006 February, 
decreasing in the MX follow-up observations of 2008 March to 7 Jy.

The second site has features in the range -1 to +13 \kms, with peak at 
+0.8 \kms\ and fairly stable at 45 Jy.  Both sites are acceptably 
interpreted as being located at  R $>$ 3.5 kpc.  
The aligned spectra clearly distinguish the emission from each site.  

\subparagraph{358.460-0.391 and 358.460-0.393}  These are both new, 
separated by 7 arcsec.  The first has a velocity range of +0.5 to +4.0 
\kms\ and its peak (at velocity  +1.2 \kms) has increased from the survey 
cube value of 25 Jy (2006 February), to the ATCA value of 35 Jy (2006 
March), and MX follow-up  (2008 March) value of 48 Jy.  The second 
maser has a peak at -7.5 \kms\ of 11 Jy which has remained stable,  and 
the velocity range of -8.5 to +6 \kms\ encompasses that of the first.  
Both are acceptably interpreted as located at R $>$ 3.5 kpc.  

\subparagraph{358.721-0.126}  We formally interpret the velocity of this 
new maser as indicating a distant site outside the solar circle, but 
with R $<$ 13.5 kpc;  however, the 
choice is not clearcut since it lies slightly outside our suggested 
boundary for this longitude-velocity domain on Fig. 4, perhaps 
indicating a location with R $<$ 3 kpc.  

\subparagraph{358.809-0.085}  The velocity of this new site 
perfectly matches the value expected for a location in the near side of 
the 3-kpc arm. 

\subparagraph{358.841-0.737}  New maser with a velocity offset from 
zero sufficient to indicate non-circular motions and a location  at R $<$ 
3 kpc.  

\subparagraph{358.906+0.106} New maser with velocity suggesting  R $<$ 3 
kpc.  

\subparagraph{358.931-0.030}  New maser with velocity similar to 
the previous site and likely to be at R $<$ 3 kpc.

\subparagraph{358.980+0.084}  New site fading from a peak of 1.6 Jy in 
the survey cube (2006 February) to less than 0.2 Jy in the MX spectrum of 
2008 March;  an  intermediate value of 0.5 Jy was measured at the 
intermediate epoch 2007 November in the ATCA measurement. The survey 
cube spectrum is shown in Fig. 1.  The velocity near zero is compatible 
with R $>$ 3.5 kpc if small non-circular motions are assumed.  

\subparagraph{359.138+0.031} For this known site, a weak feature at 
velocity -6 \kms\ was clearly visible in 1992 and 1995 with peak flux 
density of 1.0 and 0.7 Jy respectively, and although now only marginally 
detectable (0.25 Jy), was used to define the velocity range;  the 
systemic velocity is compatible with R $>$ 3.5 kpc.  

\subparagraph{359.436-0.104 and 359.436-0.102}    Both sites are 
previously known.  Their velocity ranges do not significantly overlap and 
the two sources shown on a single spectrum are distinguishable from the 
velocity ranges given in Table 1.  Both systemic velocities indicate a 
location in the near side of the 3-kpc arm.  At a distance of 5.1 kpc, 
their separation of 6.1 arcsec corresponds to 150 mpc.   
The first site is the stronger and its peak at -46.8 \kms\ has increased 
from less than 7 Jy (in 1992 and 1995) to 60 Jy in the survey cube and to 
75 Jy in the MX spectrum of 2008 March;  a feature at -52 \kms\ which was 
strongest (27 Jy) in 1992 has 
decreased to 13 Jy in 2008.  The second site has recent peak intensities 
of 1.5 and 1.6 Jy, similar to 1992 but much smaller than its value of 4.4 
Jy in 1995 (Caswell 2009).  
Forster \& Caswell (2000) show a detailed map of continuum emission 
together with information on other masers.  

\subparagraph{359.615-0.243} Known, strong  maser and highly variable.  
The peak intensity for the spectrum from the survey cube (2006 February) 
is at a velocity different from that of the follow-up MX spectrum (2008 
March), and thus variability exceeds a factor of 2; indeed, near velocity 
+22.6 \kms, intensities since 1992 have ranged from 15 Jy to 88 Jy.  The 
high variability was noted by Caswell et al. (1995b) and additional 
monitoring information was presented by Goedhart et al. (2004).  
The positive systemic velocity 
is significantly offset from zero, indicating non-circular motions and a 
location at a Galactocentric radius   R $<$ 3 kpc.  

\subparagraph{359.938+0.170} New single feature maser with flux density 
varying from 1.6 Jy in our survey cube discovery spectrum (2006 February), 
to 0.75 Jy in our ATCA measurement (2007 July) and  2.4 Jy in our MX 
follow-up (2008 March).  The near-zero velocity is compatible with  R $>$ 
3.5 kpc.  

\subparagraph{359.970-0.457}  Known maser, with one feature increasing 
from 1 Jy in 1992 to our MX measurement of 2.4 Jy in 2008 March, but the 
other fading from 0.8 Jy to less than 0.2 Jy, below our detection 
threshold.  We cite the velocity range seen in 1992, +20 to +24 \kms;  the 
velocity offset from zero indicates non-circular motions and a 
Galactocentric radius  R $<$ 3 kpc.  

\subparagraph{0.092-0.663}  New strong maser with its velocity suggesting 
R $<$ 3kpc.   

\subparagraph{0.167-0.446} New maser with intensity fading markedly 
from the survey cube value 4.4 Jy (2006 February), to 3.6 Jy during the 
ATCA measurement (2006 March) and to 1.3 Jy for the MX spectrum (2008 
March).  Its systemic velocity suggests R $<$ 3 kpc.

\subparagraph{0.212-0.001}  Known maser with velocity just within 
the range of the far side of the 3-kpc arm.  

\subparagraph{0.315-0.201 and 0.316-0.201} Known very close pair of maser 
sites,  distinguishable on the ATCA spectra shown in Caswell (1996) 
and displayed here in a single spectrum.  Intensity increases since 1995 
are less than a factor of 2.  The first site is strong with a 
wide velocity range;  the second site is weak, with its small velocity 
range of emission contained within that of the stronger site.   
The systemic velocities strongly suggest R $<$ 3 kpc and thus a 
heliocentric distance $>$ 5 kpc, for which the separation of 2.6 arcsec 
corresponds to more than 60 mpc, in accord with 
the interpretation as two separate sites (Caswell 2009).  

\subparagraph{0.376+0.040} Known maser with intensity variations from 0.7 
Jy (1992 and 1999 February), flaring to 2.3 Jy in our survey cube (2006 
February) and back to 0.62 Jy in our MX spectrum (2008 March).  Its 
velocity indicates  R $<$ 3 kpc.  

\subparagraph{0.409-0.504}  New maser, likely to be within R $<$ 3 kpc.

\subparagraph{0.475-0.010} The history of this source is given by Caswell 
(2009), and identifies the error leading to the incorrect position of 
0.393-0.034 originally reported by Caswell (1996).  The corrected position 
and its re-evaluated peak intensity of 2.9 Jy from re-analysis of the old 
data from 1995 November (Caswell 2009) agree well with our new 
measurements between 2006 and 2009.  
The velocity range as measured in the earlier data is given in Table 1; 
the systemic velocity implies a likely R $<$ 3 kpc.

\subparagraph{0.496+0.188} Known maser with many spectral features; 
the intensity is now more than twice as strong as the 10 Jy main 
peak measured in 1995 November.  The velocity 
range straddling zero velocity could be compatible with R $>$ 3.5 kpc.  

\subparagraph{0.546-0.852}  Known methanol maser site accompanied by 
water and OH masers and continuum emission (Forster \& Caswell 2000).  
Formally, the velocity suggests a distant object 
outside the solar circle, or an anomalous velocity of a site within R $<$ 
3 kpc.  However, in view of its large latitude, it has alternatively been 
interpreted as nearby (with R $>$ 3.5 kpc) and an unusually large peculiar 
motion (Gardner \& Whiteoak 1975;  Caswell 1998).  
An astrometric distance will eventually resolve this uncertainty and yield 
an excellent distance if it is indeed nearby.  Pending new data, we note 
that a far distance outside the solar circle of more than 17 kpc would 
imply an unlikely large Galactic height, z, of 250 pc.  
In contrast, a heliocentric distance between 5.5 and 8.5 kpc would not 
imply an unreasonable value of Galactic height, z, nor a contradiction to 
the apparent quite low obscuration (assuming the optical nebula RCW 142 to 
be  associated).  So  we provisionally regard it as 
an object with R $<$ 3 kpc, as was assumed by Russeil (2003).

\subparagraph{0.645-0.042 to 0.695-0.038 inclusive}  
These eleven sites in a cluster near Sgr B2 are contained within 3 arcmin 
and are individually separated by more than 10 arcsec.  Nine sites were 
seen 
in the results of Caswell (1996, 2009), and were also visible with the 
sensitivity of our present survey (although confused in the single dish 
spectrum).   The other two sites are very weak and were detected only in 
the observations of Houghton \& Whiteoak (1995) who achieved high 
sensitivity with the ATCA using a full synthesis (although with low 
spectral resolution, channel spacing 8 kHz equivalent to 0.4 \kms).   
In Table 1, for all sites, we cite the HW (Houghton \& Whiteoak 1995)
positions and velocity ranges.  We list Parkes measurements of flux 
density where they were not too confused.  We were unable to establish 
reliable peak flux density values for 0.647-0.055 (2 Jy HW; 3.4 
Jy Caswell 1996), 0.657-0.041 (1.8 Jy HW; 3.0 Jy Caswell 1996), 
0.667-0.034 (0.4 Jy HW), 
and  0.673-0.029 (0.4 Jy HW) but, for convenience, list the HW velocity  
in the Table, and the corresponding HW flux density in parenthesis after 
the reference.  Variability is difficult to establish in view of 
confusion for all measurements except those of Houghton and Whiteoak.  
Representative spectra are shown at three locations which span the 
positions of all eleven sites, with all eleven labelled to show which 
spectrum best matches their position.  Reference to Houghton \& Whiteoak 
(1995) is needed to distinguish the sites, and the labels on the 
spectra for those four sites that are extremely confused are enclosed in 
square brackets.

\begin{figure*}
 \centering
\includegraphics[width=17.5cm]{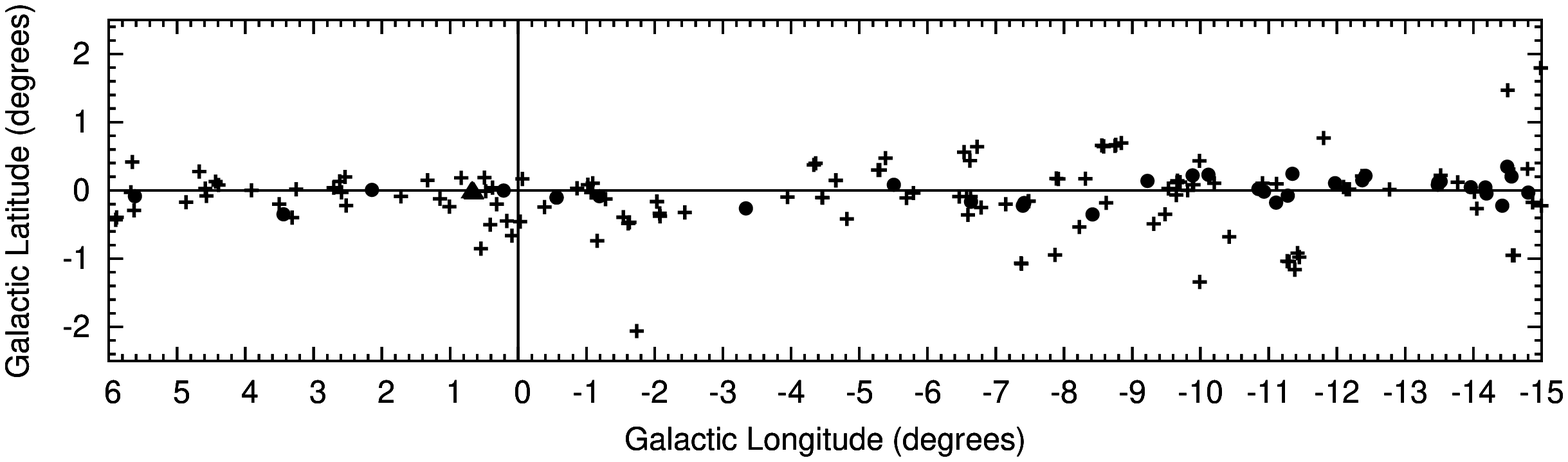}

\caption{Distribution of the 183 methanol masers in the present survey 
region.  Sources in the Sgr B2 complex are denoted by triangles, those 
believed to lie within the 3-kpc arm by circles, and the remainder 
by crosses. }

\label{fig2}

\end{figure*}

\subparagraph{0.836+0.184} Known site, with velocity close to zero  
compatible with R $>$ 3.5 kpc.

\subparagraph{1.008-0.237}  New site, with velocity close to zero  
compatible with R $>$ 3.5 kpc.

\subparagraph{1.147-0.124} New precise position (see also Walsh et 
al. 1998).  
The unusual velocity is most simply interpreted if  R $<$ 3 kpc.    

\subparagraph{1.329+0.130} New single feature maser with unusual velocity 
best interpreted if  R $<$ 3 kpc.  

\subparagraph{1.719-0.088} New site.  The velocity is slightly negative 
and requires an assumption of small non-circular motions to be compatible 
with R $>$ 3.5 kpc.  

\subparagraph{2.143+0.009}   Known maser with velocity 
well-matched to the far side of the 3-kpc arm.  There is no detectable 
uc\HII\ region  (Forster \& Caswell 2000).  

\subparagraph{2.521-0.220} New maser with velocity straddling zero 
and compatible with R $>$ 3.5 kpc.  

\subparagraph{2.536+0.198} Known site with a wide velocity range from 2 to 
20.5 \kms; the systemic velocity is compatible with R $>$ 3.5 kpc.

\subparagraph{2.591-0.029} New maser with negative velocity.  The 
assumption 
of a non-circular motion component is needed for compatibility with a 
Galactocentric radius R $>$ 3.5 kpc.

\subparagraph{2.615+0.134 and 2.703+0.040}   Two new sites 
with overlapping velocity ranges but separated by more than 7 arcmin and 
not 
confused (contrary to the similar appearance of features near +95 \kms).   
Their large positive velocities are most readily attributed to 
a location in the Galactic bar (see Section 4).  

\subparagraph{3.253+0.018} New maser with velocity straddling zero and 
compatible with R $>$ 3.5 kpc.

\subparagraph{3.312-0.399} New weak maser with small positive velocity 
compatible with R $>$ 3.5 kpc.  

\subparagraph{3.442-0.348}  New maser with single weak feature at 
a velocity well-matched to the near side of 3-kpc arm.  

\subparagraph{3.502-0.200} New maser at a quite high velocity that is not 
compatible with the 3-kpc arm nor with R $>$ 3.5 kpc;  we therefore locate 
it within R $<$ 3 kpc.

\subparagraph{4.569-0.079 and 4.866-0.171}  Two of our weakest new masers, 
with flux densities of 0.44 Jy and 0.56 Jy respectively.  

\subparagraph{5.618-0.082}  New maser with velocity well matched to the  
near side of the 3-kpc arm.  

\subparagraph{5.885-0.393} New maser with single feature at a velocity 
of +6.7 \kms.  It coincides with a previously known OH maser and strong 
\HII\ region, believed to be at a distance of 2 kpc (Stark et al. 2007).  
An earlier unsuccessful search for methanol in this direction, made in 
1992 (Caswell et al. 1995a), yielded an upper limit of 0.3 Jy, 
presumably because it was weak at that epoch.  
Our position measurement with the ATCA (2006 December) showed a peak 
intensity similar to the survey cube value (2006 August) of 1.3 Jy, but 
our MX follow-up spectrum (2008 March) showed that it had faded again, to 
0.5 Jy; both the survey cube and MX spectra are shown.  Note that 
5.885-0.0-393 is only 2 arcmin from the known source 5.900-0.430, 
and its velocity, straddled by features of the known source, suggest that 
the two sites are located at a similar distance.  We have aligned their 
spectra so as to make it clear that only a single weak feature arises from 
5.885-0.393.

\section{Discussion} 

The distribution of maser sites is displayed in Fig. 2.  
For the purposes of later discussion we distinguish sources in the Sgr B2 
complex by triangles, those believed to lie within the 3-kpc arm by 
circles and denote the remainder by crosses.  
Within 1$^{\circ}$ of the Galactic Centre, we can compare our results 
with an earlier survey (Caswell 1996) which was unbiased (not 
limited to pre-selected target positions) but was restricted to a  
narrow latitude coverage, and a small velocity range;  23 sites (11 in the 
Sgr B2 complex) were reported and another known site (0.546-0.852) lay 
outside the narrow latitude range.   
The MMB survey has been able to add only four new sites here, which 
clearly demonstrates that the full population very close to the Galactic 
Centre is indeed confined to a narrow velocity range and a small latitude 
range.   In contrast, many more new sites were found 
in the region within 6$^{\circ}$ of the Galactic Centre, but excluding the 
well-studied central 1$^{\circ}$;  here, 44 of the  60 sites listed are 
new.  The smaller number of known sources (16)  reflects the fact that 
previous searches had been limited to a modest number of pre-selected 
targets, and emphasises the value of the present sensitive unbiased 
survey.   The new full coverage of regions both sides of the Galactic 
Centre now allows us to understand, for the first time, the Galactic 
Centre population in the context of its surroundings.

\subsection{Flux densities}
The median peak flux density of the 183 sources in this part of the 
survey is 4.1 Jy.
Amongst the new detections, the only sources with peak flux density 
exceeding 20 Jy are 348.617-1.162, 351.688+0.171, 358.460-0.391 and  
0.092-0.663 (none of them exceeding 50 Jy).   
The first of these is offset by more than 1$^{\circ}$ from the Galactic 
plane, and is probably quite nearby, while the distances of the other 
three are less clear.  Seven known and seven new sources were below 0.7 Jy 
(as determined from our most sensitive MX spectra).  Four new sources were 
below 0.7 Jy in the survey discovery spectra, of which the weakest was 
4.569-0.079, which had a peak of 0.61 Jy in the discovery spectrum, and 
was subsequently measured as 0.44 Jy in the follow up MX spectrum.

\begin{figure}
 \centering
\includegraphics[width=8.5cm]{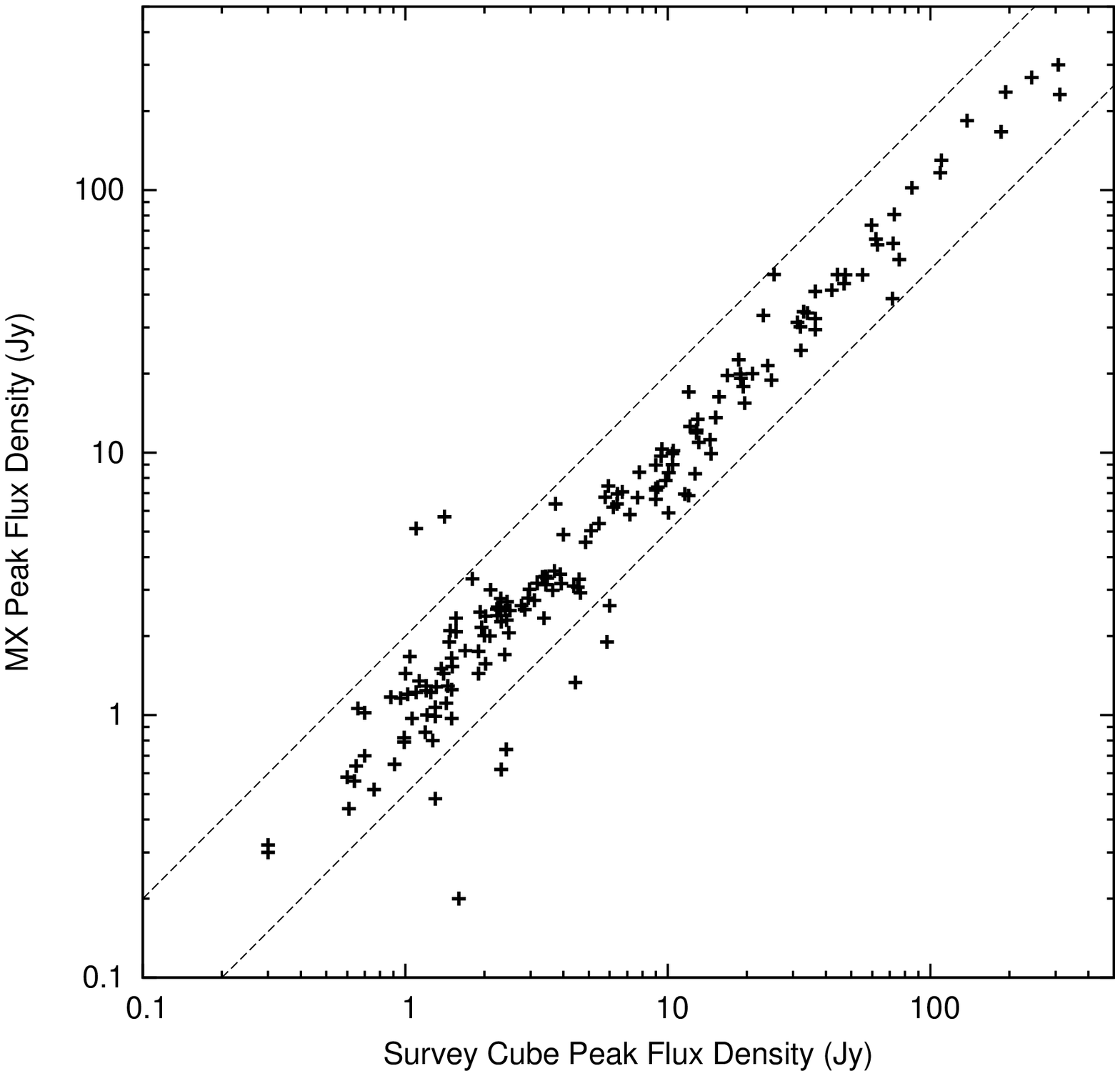}

\caption{Intensity variability revealed by comparing the Flux density 
measured during the survey with a later targeted (MX) 
measurement.  Sources within the broken lines have varied by less than a 
factor of 2.}

\label{fig18}

\end{figure}

\subsection{Variability and Completeness}

The masers in the survey have generally been observed at least three 
times.  
The peak intensities from both the survey measurement and the `MX' final 
spectrum are listed in Table 1, and a comparison at the two 
epochs is displayed in Fig. 3.  In many cases 
the differences can be attributed to small measurement errors due to noise 
in the case of weak masers, and calibration uncertainties for stronger 
sources.  The median ratio of MX peak flux density to survey cube 
peak flux density is 0.98, but some sites show significant differences 
between the two epochs, indicative of real intensity variability. The 
statistics of variability over a few epochs 
for a large number of sources was explored by Caswell, Vaile \& Ellingsen 
(1995b), and for fewer sources over many epochs by Goedhart et al. (2004).  
Our statistics are broadly compatible with the earlier extensive work, and 
when the survey is complete we will reassess the variability statistics.  
The fact that variability does occur clearly impacts on the meaning of 
`completeness' for any survey.  Our completeness in the initial survey was 
discussed by Green et al. (2009a) and estimated to be 80 per cent at the 
0.8 Jy level, and a progressively smaller percentage detected down to 0.6 
Jy.  The criterion for inclusion in the catalogue is that a high precision 
position from the ATCA must have been obtained.  
Our preliminary catalogue contained a few weak apparent detections that we 
were unable to confirm with the ATCA.  Since they were 
close to our detection limit, we cannot distinguish whether they varied, 
or were spurious, and we omit them from the final catalogue.  
For most of the ATCA observations, a target would have been detectable 
even if it had faded to one-quarter of its initial intensity.  We drew 
attention in the notes of Section 3 to nine sources that varied by more 
than a factor of two during our observations.  This is a small fraction (5 
per cent) of the survey total and so we expect very few of the 
sources not confirmed by the ATCA to be real ones that faded.  
Observations of the `piggyback survey' (described by Green et al. 2009a) 
are approximately four times more sensitive than the main MMB survey and 
cover a significant fraction of the main survey region.  They are 
currently being processed and will result in some new sources below the 
sensitivity limit of the present survey (the weak example 
`piggyback' source displayed by Green et al. (2009a) is at a position  
covered by the present part of the survey).  We envisage publishing 
`piggyback sources' as a supplementary weak source catalogue, and will 
include any sources from the main survey which faded before the ATCA 
observations, but reappeared during future checks.  Comparison of the 
piggyback results with the main MMB survey will also be used to more 
rigorously assess the completeness of the main survey.  

There are nine sources where the peak has varied by more than a 
factor of two in our observing sessions since 2006 (see Fig. 3).  For 
such strong variations it should be especially simple to discover from 
future monitoring whether any of these show strict periodicity such as has 
been established for seven enigmatic methanol masers to date (Goedhart et 
al. 2004 and citations thereto).  
In one instance (the new source 353.216-0.249) we have seen a rise, from 
1.1 Jy to  a peak of 5.1 Jy (MX on 2008 March 18) followed by decline to 
less than 1 Jy, when we made additional measurements (2009 April 5).  
359.938+0.170 showed a drop from 1.6 Jy at its discovery in our survey 
cube to 0.75 Jy in our ATCA measurement followed by a recovery to 2.4 Jy 
in our MX follow-up 2008 March.
The other seven strong variations during our observations were 
an increase for 345.985-0.020 and marked decreases for 349.579-0.679, 
351.242+0.670, 358.980+0.084, 0.167-0.446, 0.376+0.040 and 5.885-0.393.

At least 16 previously known sources in our survey region have 
a history of variability, or show newly discovered variability when 
compared with our new measurements:
346.517+0.117, 346.522+0.085, 347.863+0.019, 
350.011-1.342, 350.105+0.083, 350.116+0.084, 
352.083+0.167, 353.273+0.641, 354.615+0.472, 358.386-0.483, 
359.436-0.102, 359.436-0.104, 359.615-0.243, 359.970-0.457, 0.376+0.040 
and 0.496+0.188.
For six of these, the large changes are clearly seen from comparison of 
the spectra shown in this paper with Parkes spectra taken 15 years 
earlier, in 1992 (Caswell et al. 1995a).  For some of the 
others, current spectra are comparable with those of 1992 and there was a 
flare or decline at an intermediate epoch (Caswell 2009) as detailed in 
the source remarks of Section 3.1.

More generally, there is also a wide range of variability in subsidiary 
features of these and many other masers, which will be the subject of 
future studies.

\subsection{Galactic latitude distribution}

The Galactic latitude distribution can be seen from Fig. 2.  
Our surveying strategy readily allowed extension to a larger latitude 
range than $\mid b \mid \le 2^{\circ}$ if the detected distribution had  
indicated that it might be worthwhile.  This proved unnecessary even in 
the Galactic Bulge region, since our inspection of the latitude 
distribution 
of the 183 detected sources shows a tight concentration to low latitudes, 
with 173 lying within 1$^{\circ}$ of the Galactic plane.  The 10 
sources outside this latitude range (and within the currently presented 
longitude range) are:
345.010+1.792 with 345.012+1.797, 345.498+1.467, 348.617-1.162, 
348.703-1.043 with 348.727-1.037, 350.011-1.342, 352.630-1.067 
with 352.624-1.077, and 358.263-2.061.  

Of these, only 348.617-1.162 is a new detection, and only 358.263-2.061 
lies outside our regular survey coverage.  

Overall, it is clear that towards the Galactic Centre, there is no 
significantly broader latitude distribution than elsewhere, and thus 
the Galactic Bulge that is recognisable from other varieties of objects 
contains no significant population of current high mass star formation 
regions.

From a pragmatic view, we argue that the narrow latitude distribution over 
all longitudes indicates that very few sources above our detection 
limit lie undiscovered more than 2$^{\circ}$ from latitude zero.  

The significant interpretation from the narrow latitude distribution is 
that the masers are confined to a very thin disk and the 
maser population is dominated by distant objects beyond 
a few kpc.  This will be discussed in more detail when the remainder of 
the survey is presented.

\subsection{Absence of very high velocities}

We recall that our velocity coverage of the region surveyed here, spanning 
the Galactic Centre, was very large, with the objective of covering all 
velocities where CO emission had been detected.  Our maser detections are 
all found to lie within the quite small velocity range -127 to +104 \kms.  
Within 2 
degrees of the Galactic Centre, no detections were made outside the even 
smaller range of only -60 to +77 \kms.   
In sharp contrast to the masers, the CO in this region shows a wide 
velocity range which extends from -260 to +280 \kms (Dame et al. 2001), 
commonly interpreted 
as material in rapid rotation extremely close to the Galactic Centre.   
The absence of accompanying high velocity masers is an interesting result 
from our survey and may indicate that the conditions in this region are 
hostile to massive star formation (see also Section 4.6.3).  
Alternatively, the absence of detected masers from this molecular material 
may simply reflect the fact that the region considered is a very small 
volume, with very few masers expected unless their space density were 
extremely high.

\subsection{Velocity ranges}

As noted in Section 2, the velocity range (lowest and highest velocities) 
that we have chosen to list for each source in Table 1 corresponds to 
extreme values seen at any epoch, since this range is likely to be more 
meaningful in characterising the site than the individual values at 
different epochs, which depend mostly on variability and observational 
sensitivity.  Only two of the 183 masers have ranges greater than 16 
\kms.    
These are the new source 353.429-0.090 with range nearly 19 \kms; and 
the known source 2.536-0.198 with range 18 \kms.  The range distribution 
is similar to that of the larger sample of Caswell (2009) which 
showed velocity widths exceeding 16 \kms\ to be rare, accounting for a few 
per cent of the sources.  
The mid-value of the range is our preferred estimate for the systemic 
velocity of each site, a preference justified by Szymczak, Bartkiewicz \& 
Richards (2007) and Caswell (2009).

\begin{figure}
 \centering
\includegraphics[width=8.4cm]{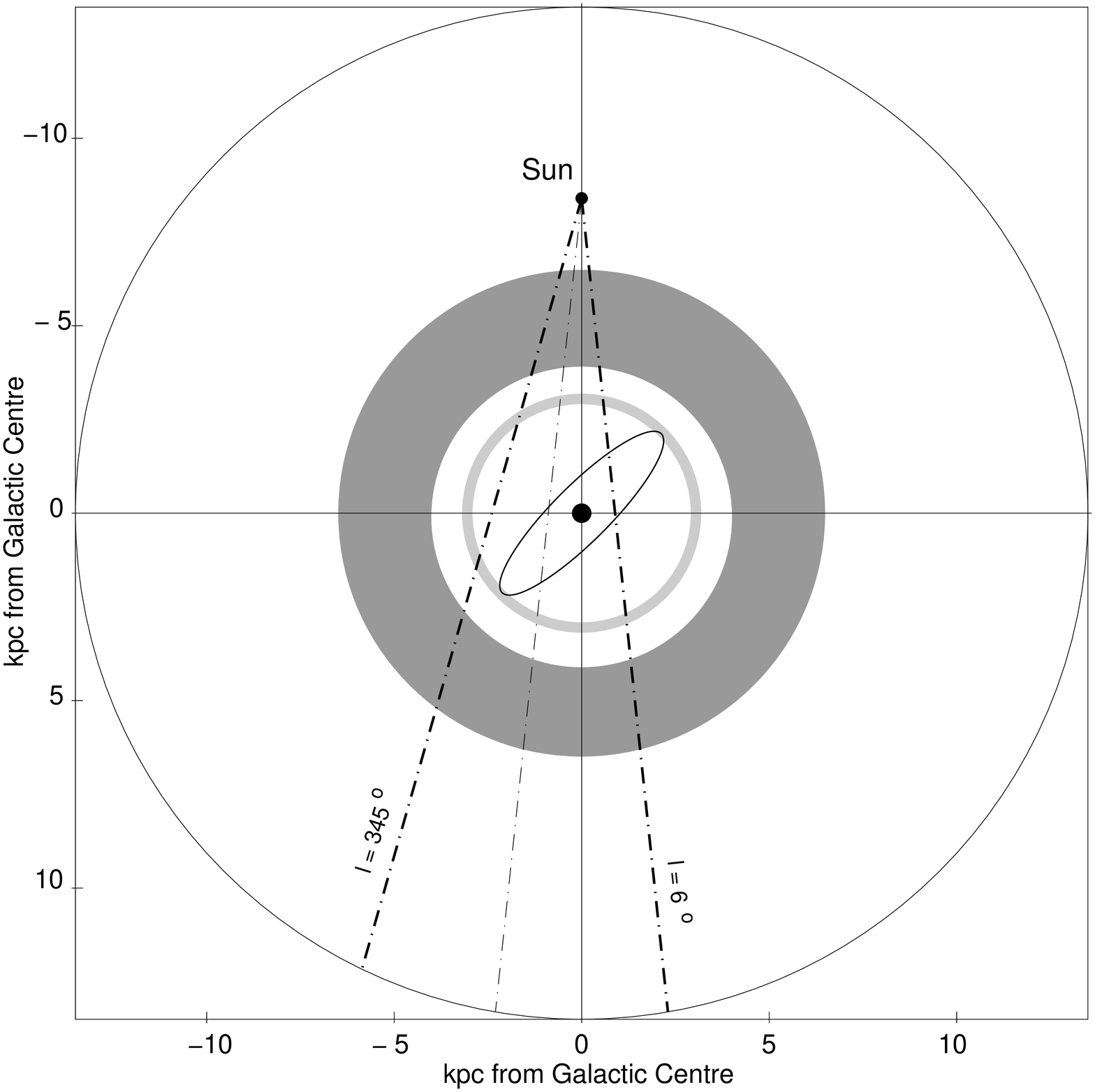}
\caption{Plan view of the Galaxy extending to the outer boundary of 
detected methanol masers (Galactocentric radius 13.5 kpc). 
A central dot, radius 0.25 kpc, denotes the Galactic Centre Zone,  a 
lightly shaded annulus the 3-kpc expanding arm and a shaded 
annulus from radius 4 to 6.5 kpc shows where high mass star formation is 
concentrated.  An 
ellipse depicts the likely location of a Galactic bar.  Labelled 
longitudes $345^{\circ}$ and $6^{\circ}$ mark the limits of the 
Galactic sector covered in the present survey.  }

\label{fig19}

\end{figure}

\subsection{The space density and kinematics of masers}

Our main aim in this section is to compare the Galactic Centre 
region with the remainder of the Galaxy, particularly with regard to the 
space density of maser sites which, in turn, can be interpreted as 
revealing the relative current rates of high mass star formation.  
The relevance of our results can be seen in Fig. 4 showing the 
coverage of the current portion of the survey in relation to the whole 
Galaxy.  
At large Galactocentric radius, R, there are vey few masers and we 
therefore limited  the outer boundary to R = 13.5 kpc, beyond which there 
are no authenticated methanol maser sites and very 
little high mass star formation.  
The Galactocentric radius of the Sun, R\subsol, is depicted at 8.4 kpc 
(see later).  We also mark several distinct zones 
that will be discussed in detail later.  These are:  
an annulus from 4 to 6.5 kpc which is known to have a high density of 
molecular material hosting massive star-forming sites and masers;  
an annulus from R =  3 to 3.5 kpc corresponding to the 3-kpc arm;  
the approximate extent of a Galactic bar; 
and a small region at the centre with radius 250 pc which we refer to as 
the Galactic Centre Zone; 
We note that maser sites physically near the Galactic Centre can 
be recognised firstly by their confinement to a small 
longitude range, but require distance estimates to distinguish them from 
foreground and background sites.

\begin{figure}
 \centering
\includegraphics[width=8.1cm]{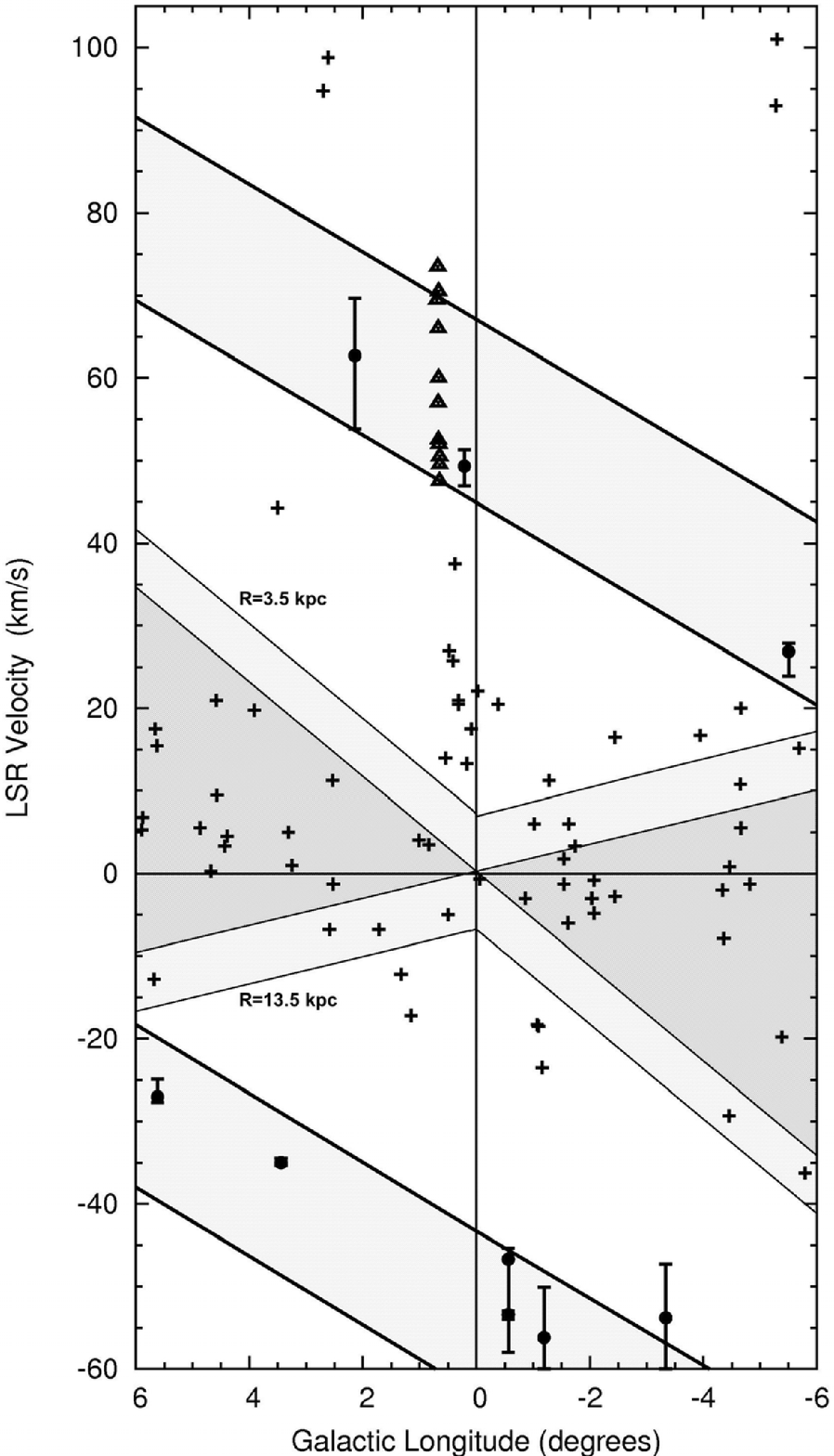}
\caption{Longitude-Velocity plot for masers within 6 degrees of the 
Galactic Centre direction.  Maser site symbols are crosses except for 
those within 
Sgr B2 (triangles) and nine sites attributed to the 3-kpc arm (filled 
circles with velocity range bars).  The central wedge-shaped shaded areas 
are the l-v domains for sites with Galactocentric radii between 3.5 and 
13.5 kpc, bounded by lighter shading to allow for small deviations from 
circular rotation (see text).   The wide lightly-shaded parallel sloping
bands near the bottom and top are the domains of the near and 
far sides of the 3-kpc arm.  Sites within the remaining white domain are 
believed to lie within 3 kpc of the Galactic Centre, as 
discussed in the text.   } 

\label{lv6plot} 

\end{figure}

Fortunately, the unique kinematics near the Galactic Centre can act as a 
distance discriminator, as we will see from a discussion of Fig. 5, which 
shows the longitude versus velocity distribution for maser sites  with  
$\mid l \mid \le 6^{\circ}$.    We have indicated several key 
domains in longitude-velocity (l-v) space.  The precise location of the 
lines drawn depends on the Galactic rotation curve chosen (see next 
section), but the overall appearance is not affected by that choice.  We 
will be emphasising the inferred Galactocentric radius 
of each maser site and will be less interested in the heliocentric 
distance.  Objects at the Galactocentric radius of the Sun, R\subsol,  
lie on the zero velocity locus (with 
minor variations if changes are made to the adopted solar motion).  We 
bracket this region with the loci of objects in strict circular rotation 
at R values 3.5 and 13.5 kpc, so objects with 3.5 kpc $<$ R $<$ 13.5 kpc 
will lie in the shaded central region.  We have increased the boundaries 
of this region 
by 7 \kms\ to allow both for small deviations from circular rotation and 
for uncertainties in determining the systemic velocity for individual 
objects.  The lower boundary of 3.5 kpc was chosen 
because extreme departures from circular rotation are known to be present 
inside this region.  The most notable structure within 3.5 kpc is the 
`3-kpc arm'.  We retain this common designation despite the 
fact that it is neither a simple spiral arm, nor is 3 kpc the best 
estimate of its Galactocentric radius.  
It is usually regarded as a ring, with R approximately 3.3 kpc and its 
most striking feature is an expansion outwards from the Galactic Centre at 
approximately 50 \kms\ (see next sections).   The wide lightly-shaded 
parallel sloping bands near the bottom and top of Fig. 5 correspond to 
the domain of the 
near and far sides of this ring, with 
parameters suggested by Dame \& Thaddeus (2008) and as applied to an 
initial investigation of the present survey by Green et al. (2009b).  
Note that some prominent sites lying in this domain, members of 
the Sgr B2 complex, are not part of the 3-kpc arm, but lie much closer to 
the Galactic Centre.  

Fig. 5 highlights the fact that near Galactic longitude zero there is 
very little velocity discrimination for sources over the large range of R 
between 3.5 and 13.5 kpc, where circular rotation dominates.  This lack 
of discrimination is usually regarded as unfortunate and preventing the 
assignment of meaningful kinematic distances close to Galactic longitude 
zero.  However, for our special purpose of recognising the Galactic Centre 
population, it is a strength rather than a weakness.   With reference to 
Fig. 5, if we accept the premise that essentially all maser sites with R 
$>$ 3.5 kpc can lie only in the central l-v domain (dark shading with 
lightly shaded error band), and that the Galactocentric annulus 
between 3 and 3.5 kpc radius is wholly occupied by the 3-kpc ring sources 
(lightly shaded parallel bands with large offset from zero velocity), 
then sources in the unshaded l-v domain must all arise within 3 kpc 
of the Galactic Centre where non-circular motions can dominate.  The 
number in the unshaded area  will be a lower limit to (conservative 
estimate of) the number with R $<$ 3 kpc since some can 
share the same central l-v domain as the sources with R $>$ 3.5 kpc.

\subsubsection{Galactocentric radii $>$ 3.5 kpc}

We consider this region first so as to establish an estimate for the 
maximum space density of masers in the spiral arms, at Galactocentric 
radii $>$ 3.5 kpc, where high mass star formation chiefly occurs.  
This can then be a yardstick for comparison with the Galactic Centre.  
The small portion of the MMB survey presented in this paper is not 
suitable for detailed studies at R $>$ 3.5 kpc.  However, the whole MMB 
survey southern region 
from Galactic longitude 186$^{\circ}$ through the Galactic Centre to 
60$^{\circ}$ (as described by Green et al. 2009a) is essentially complete 
and, even in preliminary unpublished form, can provide an estimate 
adequate for our purpose.   The study of spatial 
distributions also requires distance estimates for each source.    
Accurate distances are now being derived for methanol masers from VLBI 
measurements sufficiently precise to allow parallax determinations, but 
only a handful of sites have  so far been measured (Reid et al. 2009).  We 
must therefore rely at present 
on kinematic distances for most sources.  However, the existing 
astrometric distances have already revealed a number of insights which 
suggest a slightly improved prescription for determination of kinematic 
distance estimates (Reid et al. 2009).  Reid et al. adopt a flat rotation 
curve (with velocity 254 \kms) and incorporate a significant revision to 
the standard solar motion currently used to adjust velocities to the LSR   
(see also McMillan \& Binney 2009;  Binney 2009); they also apply a 
velocity correction (which partially counteracts the solar motion 
revision) assumed to be applicable to high mass young stars.  This latter 
correction is somewhat more controversial, possibly indicative of the need 
for further revision to the solar motion (McMillan \& Binney 2009, 
Binney 2009).  
For our purposes, the interpretation of the correction is not important 
since it does not greatly affect the resulting distances.  
We note that Reid et al. (2009) incorporate a Galactic Centre 
distance R\subsol\ of 8.4 kpc, rather than the IAU recommended value of 
8.5 kpc.  This scaling change is too small to noticeably 
affect the commonly described features within 3.5 kpc of the Galactic 
Centre, so we retain, for example, the mean radius 
for the 3-kpc arm of 3.3 kpc suggested by Dame \& Thaddeus (2008).  

We provisionally applied the Reid et al. (2009) prescription for kinematic 
distances to our preliminary unpublished MMB 
catalogue to derive the space density of masers as a function 
of Galactocentric radius R, for R $>$ 3.5 kpc.  Note that for this purpose 
there is no need to discriminate between near and far distance 
ambiguities.   A detailed analysis will be deferred until the full 
catalogue is finalized, but our initial studies (with no allowance for 
distance and luminosity 
effects) clearly show a broad peak in the radial distribution between 
Galactocentic radii 4 and 6.5 kpc at least as high as 7 masers per 
kpc$^2$, where the area refers to the area in the plane of the Galaxy.  
This peak is an azimuthal average.  It thus corresponds to a value smeared 
over spiral arm and inter-arm regions, so the space density within 
the arms is likely to be at least three times as high, i.e. 20 masers per
kpc$^2$.   A peak between 4 and 6 kpc was also evident in the smaller 
sample of methanol masers (with poorer, and inhomogeneous, sensitivity 
limits) studied by Pestalozzi et al. (2007).
The Galactic distribution of OH masers shows a similar peak between radii 
4.25 and 6 kpc (Caswell and Haynes 1987; after scaling from their 
assumed Galactic Centre distance of 10 kpc to 8.4 kpc which is the only 
major change needed for a valid comparison with our present results).

\subsubsection{The 3-kpc arm (expanding ring)}

We now turn to the inner region of the Galaxy with R $<$ 3.5 kpc.  Here, 
the atomic and molecular gas has long been known to exhibit unusual 
kinematics (e.g. Cohen \& Davies 1976 and citations thereto), and features  
which are still not well understood.  Most notable is the expanding 
3-kpc arm, for which we adopt the parameters suggested by Dame \& Thaddeus 
(2008).  They treat it as a ring or annulus with mean radius of 3.3 kpc 
(based on the estimated location of tangent points to the ring as 
recognised in HI and CO at longitude $\pm$23$^{\circ}$), and 
showing an expansion velocity of between 53 and 56 \kms, over and above 
circular rotation.    

In the longitude velocity domain, the Dame \& Thaddeus (2008) CO data 
allow clear recognition that the 3-kpc arm features lie on lines with a 
mean slope of  4.12 \kms\ per degree.  For a ring at Galactocentric radius 
R, this slope (strictly a sin(longitude) function, but close to a linear 
slope for small longitude) is very simply related to R, and its rotation 
velocity  (e.g. Cohen \& Davies 1976).  
If we adopt a value of R = 3.3 kpc, with R\subsol\ = 8.4 kpc and rotation 
velocity 254 \kms\ at the Sun, the slope of 4.12 \kms\ per degree implies 
a rotation velocity at 3.3 kpc of 193 \kms, i.e. lower than at the 
Sun by 61 \kms\ (more than 24 per cent).  
This is clear evidence that it would not be appropriate to assume a flat 
rotation curve extending from the solar circle to R as small as 3.3 kpc.

The distinctive kinematic signature of the 3-kpc ring (see Fig. 5), 
interpretable as a well-defined rotation velocity modified by an apparent 
expansion of more than 50 \kms, has allowed the 
recognition of many methanol masers that lie within it (Green et al. 
2009b).  The presence of the masers demonstrates the ongoing massive star 
forming activity in the 3-kpc arms, which had earlier been in dispute.  
In the source notes, we identify the individual sources that are likely to 
lie in the 3-kpc arm, after making some small revisions to the 
preliminary assessment by Green et al. (2009b).  We also draw attention to 
the presumed `start' of the far 3-kpc arm near longitude 
346$^{\circ}$ where it is assumed to originate, at the end of a Galactic 
bar.  This appears to be a densely populated portion of the far 
arm, and while some outer spiral arm confusing sources may also be 
present in this domain of longitude-velocity space, the local over-density 
may well be an indication of especially active star formation in this 
unique part of the 3-kpc arm.  
We estimate that the mean space density of maser sites in the ring is 
approximately 4 per kpc$^2$, when we approximate it to an annulus between 
Galactocentric radius 3 and 3.5 kpc, but make allowance for the limited 
domain where the kinematic signature is distinct from the outer spiral 
arms.

\subsubsection{Region interior to the 3-kpc arm}

Here we will distinguish between two discrete spatially confined 
structures and a more diffuse population within the inner 3-kpc of the 
Galaxy.  The discrete structures are the Galactic bar and a small 
Galactic Centre Zone (GCZ) which we define as within a radius of 250 pc of 
the Galactic Centre (see Fig. 5).  

Past discussions have recognised this Galactic bar, and a loosely 
defined central molecular zone (e.g. Morris \& Serabyn 1996), both of 
which have already been noted as containing methanol masers (Caswell 1996, 
1997). Our GCZ is a more precisely defined region which encompasses the 
central molecular zone.  

With regard to the Galactic bar, the previously known maser 354.724+0.300 
has already been attributed to this structure (Caswell 1997).  It has a 
velocity that, for this longitude, is extremely anomalous (near +100 
\kms);  it is 
loosely associated with an HII region postulated to lie within the 
elliptical orbits of the Galactic bar (Caswell \& Haynes 1982).  The 
present survey has discovered a nearby companion maser 354.701+0.299.  
Two other masers, 2.615+0.134 and its companion 2.703+0.040, also have 
unusual velocities, near +94 \kms. They could probably also be 
accommodated within the bar  (far side of the ellipse shown in Fig. 3, 
and in fig. 1 of Caswell \& Haynes 1982), and we regard this as the most 
plausible explanation.

The  GCZ includes Sgr B2 as a distinct complex within it.  This fact 
alone indicates that there is significant massive star forming activity 
in the GCZ but our objective is to assess whether the GCZ shows enhanced 
star forming activity additional to the Sgr B2 complex.   From our 
definition of the GCZ as a disc of radius 250 pc centred on the nucleus, 
we note that for a Galactic Centre distance of 8.4 (or 8.5) kpc,  the only 
maser sites that can be attributed to the GCZ must lie 
within the longitude range   358.3$^{\circ}$ to 1.7$^{\circ}$.  
This longitude sector will also include foreground and background 
sources.  We can crudely estimate the foreground and background sources by 
making a comparison with the longitude range 
1.7$^{\circ}$$<$ $\mid l \mid$ $<$ 3.4$^{\circ}$,
which statistically is expected to have similar numbers of foreground and 
background sources but cannot have any in the GCZ.  
We find 17 masers in the comparison longitude range, compared with 41 in 
the range $\mid l \mid$ $<$ 1.7$^{\circ}$.  This suggests that 24 sites 
(including 11 within the Sgr B2 complex) lie in the GCZ.  A similar result 
is obtained by extending the comparison region by a further 1.7$^{\circ}$.  
Thus statistically we have confirmed an earlier conclusion (Caswell 
1996) that there is indeed a clear population of masers in the GCZ. 
However this simplistic approach does not make use of the 
valuable velocity information, and we now attempt to make the estimate 
more precisely, and identify the location of individual maser sites. 

We first require an estimate of the mean space density of smoothly 
distributed masers within Galactocentric radius 3 kpc (which are 
necessarily confined  within 20.9$^{\circ}$ of the Centre).  
With the exclusion of the 4 Galactic bar masers and those that could  
lie within the GCZ, we have estimated this to be 1.8 sources 
per kpc$^{2}$.  This value was derived using the preliminary MMB 
catalogue, from consideration of the longitude 
range within 20.9$^{\circ}$ of the Centre, but with the exclusion of the 
small sector 358.3$^{\circ}$ to 1.7$^{\circ}$, and removing sources 
that lie within the 3-kpc arm and those with kinematic Galactocentric 
distances greater than 3.5 kpc.  

We now return to the 41 masers in the small sector within 1.7$^{\circ}$ of 
the Centre.   We discuss these in our notes of Section 3.1 where we remark 
on the likely location of each of them based on their velocities and 
other available evidence.  From these individual considerations and the 
overall statistics, we suggest that 10 of the 41 masers may lie more than 
3.5 kpc from the Galactic Centre, these being the masers with LSR 
velocities close to zero (see Fig. 5).  4 sites are attributed to the 
3-kpc arm.  This then leaves 27 sites that appear to lie 
within 3 kpc of the Galactic Centre.   22 of these are expected to  
lie within the GCZ, since 5 possibly lie in the smoothly 
distributed population at radii between 250 pc and 3 kpc (using our 
estimated space density of 1.8 per kpc$^{2}$).  These assessments will 
eventually be verifiable from precise astrometry.   

The 22  sites  attributed to the GCZ include 11 within the Sgr B2 
complex. The area in the plane of the Galaxy of the GCZ is 0.2 kpc$^{2}$ 
and thus apart from Sgr B2, the space density of masers in the GCZ is 
55 per kpc$^{2}$;   the 11 Sgr B2 sites increase it to 
110 per kpc$^{2}$.  The space density is thus more that 50 times higher 
than in the immediate surroundings, and also somewhat higher than 
the average in the spiral arms of about 20 per kpc$^{2}$ (Section 4.6.1).  
However, as noted by Caswell (1996), for such small regions, and taking 
into account the short lives of massive stars, the 
statistical fluctuations over time will be large, and there are likely to 
be similarly small regions in spiral arms with comparably high maser 
space densities at some epochs, as will become clear when statistics 
from the final catalogue are available.  Our present conclusion is that 
the GCZ has a high current rate of star forming activity, rivalling, and 
probably exceeding that in the spiral arms.  It reinforces and extends 
a conclusion (Caswell 1996) based on simple considerations of an earlier  
dataset restricted to a very small longitude range near the Galactic 
Centre.

\subsection{Data availability and survey follow-up}

All survey data sets will be made available online in due course, and will 
be a resource enabling archival searches for methanol emission (or 
absorption) towards future targets of interest. 
Global investigations of the maser distribution from the methanol 
survey and investigations of related quantities (such as the 
luminosity function) will be presented when final results for more of the 
Galaxy have been completed.  

Comparisons with CO and HI (in some cases allowing 
distance discrimination between near and far kinematic distances) and 
other surveys are in progress and results will be deferred until a larger 
portion of the catalogue has been completed.  Discussion of correlations 
with other objects such as those in the IR will also be deferred.

In addition to correlations with published data, we are conducting a 
number of new follow-up observational programs targeted towards the maser 
sites, in order to study various other  methanol transitions, and also 
other molecular masing species.    

The other methanol transitions include not only other masing transitions 
that are of the class II variety such as 12 GHz, but also Class 
I methanol masers such as those at 36 and 44 GHz.

Foremost amongst masers of other molecular species is the 6035-MHz 
transition of OH,  which was surveyed in parallel with the methanol (Green 
et al. 2009a) but with results to be presented as a separate study.
Other studies are being conducted for associated ground-state OH and for 
water masers at 22 GHz.  The high frequency maser follow-ups will include 
studies of the primarily thermal emission from other molecular species 
such as ammonia and SiO, and deeper studies of the continuum to 
complement our existing 8.6-GHz data in searching for ultracompact and 
hypercompact \HII\ regions associated with the masers or in their 
vicinity.

\section{Conclusions} 

This first portion of our Galactic Plane survey of methanol masers 
at 6668 MHz has emphatically confirmed the benefits of a survey which is 
unbiased rather than limited to pre-selected targets, and has generously 
covered ranges of velocity and Galactic latitude beyond pre-conceived 
expectations of where the masers would be concentrated.  We have 
corroborated the confinement of maser sites to a thin disk population, 
even in the Galactic Centre region.  Specifically, masers close to 
the Galactic Centre show no recognisable high latitude population that 
might be ascribed to the Galactic Bulge, and there are no sites at extreme 
velocities.  
However, there is a clear population associated with the `3-kpc arm', both 
the near and far portions.
There is also a high density of masers in the Galactic Centre zone, 
within 250 pc of the Galactic Centre.  We have demonstrated this 
statistically, and identified the likely members from their kinematics.
Results for the whole Southern Galaxy are nearly complete and will be 
presented soon, allowing more detailed comparisons of the Galactic Centre 
with the remainder of the Galaxy.

\section*{Acknowledgments} AA and DW-McS acknowledge support from a 
Science and Technology Facilities Council (STFC) studentship. LQ
acknowledges support from the EU Framework 6 Marie Curie Early Stage
Training programme under contract MEST-CT-2005-19669 ``ESTRELA''.  
The authors thank staff at the Parkes Observatory and the Australia 
Telescope Compact Array for ensuring the smooth running of the 
observations.  Both facilities are part of the Australia Telescope which 
is funded by the Commonwealth of Australia for operation as a National 
Facility managed by CSIRO. 

We dedicate this paper to the memory of our friend and colleague Jim 
Cohen in recognition of his immense contribution to this project from its 
inception;  the Galactic Centre focus in this first survey 
paper is especially appropriate in view of Jim's major contribution to 
early studies of the Galactic Centre and the interpretation of its HI 
kinematics.

\bsp
\label{lastpage}
\end{document}